\documentclass[11pt,a4paper]{article}

\usepackage[a4paper]{geometry}

\usepackage[
    style=alphabetic,   % cite as [Wit89, CY19]
    backref=true        % bibitems link back to reference
]{biblatex}
\usepackage{pgfplots}   % also loads TikZ which loads xcolor
\pgfplotsset{compat=1.18}

\usepackage{mathtools}  % also loads amsmath

\usepackage{amssymb}    % also loads amsfonts

\usepackage{hyperref}
\hypersetup{
    colorlinks=true,    % colour the text of links
    linktocpage=true    % link in toc is page number, not text
}

\usepackage{subcaption}

\usepackage{LTC}

\title{Higher-Dimensional Integrable Scattering}
\author{Lewis~T.~Cole}

\begin{document}

\begin{titlepage}
\centering

\vspace*{2cm}
{\Large\bfseries
Higher-Dimensional Integrable Scattering}\par
\vspace{1.2cm}

{\large
Lewis~T.~Cole}\par
\vspace{0.8cm}
\textit{\normalsize
School of Mathematics and Maxwell Institute for Mathematical Sciences,\\
University of Edinburgh, EH9 3FD, UK}\par
\vspace{0.6cm}
{\normalsize
\texttt{lewis.cole@ed.ac.uk}}\par
\vspace{2cm}

\textbf{Abstract}\par
\vspace{0.4cm}
\begin{minipage}{0.85\textwidth}
\small
\noindent
This paper describes classical soliton scattering in the 2+1d integrable chiral model,
related to the Bogomolny monopole equation and self-dual Yang-Mills,
as a case study for higher-dimensional integrable scattering.
Extended line solitons exhibit nontrivial scattering
with lump solitons and with other line solitons,
and $N \to N$ interactions can be factorised into a series of $2 \to 2$ interactions.
Scattering between parallel line solitons in 3d is related to lump soliton scattering in 2d,
allowing for direct comparison with results in 1+1d integrable field theories.
\end{minipage}
\vfill

\end{titlepage}

\newpage
\hrule
\tableofcontents
\vspace{2em}
\hrule
\vspace{1em}

\section{Introduction}
Integrable systems play a valuable role in theoretical and mathematical physics
because many of them can be solved exactly.
For example, all instantons in Yang-Mills can be constructed by solving
the integrable self-dual Yang-Mills equation~\cite{ADHM78},
and the scattering amplitudes of certain 1+1d integrable quantum field theories
can be found to all orders via an integrability bootstrap program~\cite{ZZ79}.
While there is no universal definition of integrability,
integrable nonlinear PDEs can often be reformulated in terms of auxiliary linear problems,
and integrable quantum field theories typically have
infinitely many symmetries and conserved charges.

\medskip

In more than two spacetime dimensions,
integrable quantum field theories have been said to exhibit trivial scattering
as a consequence of the Coleman-Mandula no-go theorem~\cite{CM67}.
The argument can be presented heuristically as follows.
Integrable field theories have higher-spin symmetries which shift the trajectories of particles
by an amount that depends on their momentum.
Starting with a configuration in which two trajectories meet at some point in spacetime,
these symmetries will produce a configuration in which the two trajectories miss one another,
meaning that both configurations must describe a trivial scattering process.
Quantum field theories in 1+1d evade this argument by the simple observation that
two lines generically intersect somewhere in the plane.

\medskip

However, there are at least two ways to get around this argument in higher dimensions.
First, instead of scattering particles with generic momenta,
we can impose kinematic constraints which relate the momenta to one another.
In the 2+1d integrable chiral model (which will be the focus of this paper),
this leads to nontrivial classical soliton scattering,
observed numerically in~\cite{Sut92} and then computed analytically in~\cite{War95}.
\unskip\footnote{
The kinematic constraint corresponds to a limit
in which marked points on $\CP^1$ coincide~\cite{War95}.
Further examples with two coincident points were studied in~\cite{Ioa96},
and the limiting procedure was generalised to an arbitrary number of points in~\cite{DT07}.
}
More recently, tree-level scattering amplitudes of 2+2d self-dual Yang-Mills
and 2+2d self-dual gravity (both integrable theories) have been shown to be nonvanishing
under certain kinematic constraints~\cite{GLSSW26a,GLSSW26b}.

\medskip

Second, we can scatter extended objects whose trajectories
will generically intersect at some point in spacetime.
Solitons in integrable field theories are exact solutions to the classical equations of motion,
and in higher dimensions their energy densities
can be localised around submanifolds of various dimensions.
We will refer to them as lump solitons or line solitons
depending on whether their energy densities are localised around points or lines in space.
Line solitons in self-dual Yang-Mills have been shown to exhibit nontrivial scattering~\cite{Veg88},
and in the 2+1d integrable chiral model, line solitons interact nontrivially
with lump solitons and with other line solitons~\cite{Lee89}.
More recently, interactions between line solitons in the 2+2d Wess-Zumino-Witten model
have been studied in~\cite{LHHZ25}.
This second option, concerning the scattering of extended objects,
will be the central topic of this paper.

\medskip

The purpose of this paper is to describe classical soliton scattering
in the 2+1d integrable chiral model as a case study for higher-dimensional integrable scattering.
In \secref{sec:3dICM}, we review the basic features of the 2+1d integrable chiral model,
including its relation to various other theories.
In \secref{sec:solitons}, we review the solution generating technique known as the dressing method,
and use it to construct soliton solutions to the 2+1d integrable chiral model.
This includes lump solitons which have finite total energy,
and line solitons which can be interpreted as domain walls
interpolating between two different vacua.
Many of these solutions are known, but some appear to be new.

\medskip

In \secref{sec:scattering}, we study the classical scattering of these solitons
at the level of the exact field configurations.
The dressing method allows us to construct solutions to the nonlinear equations of motion
which asymptotically look like the superposition of multiple isolated solitons.
By comparing the exact field configurations in the far past and in the far future,
we can derive a classical notion of scattering for these solitons.
For $2 \to 2$ interactions, we derive a general expression for the transformations of the solitons
which reproduces many of the known results in the literature.
Furthermore, we describe a factorisation of $3 \to 3$ interactions into $2 \to 2$ interactions,
and demonstrate that this factorisation is consistent
because it does not depend on the order of interactions.
We discuss but do not fully establish a connection between these statements
and the 2d Yang-Baxter equation and the 3d Zamolodchikov tetrahedron equation.

\medskip\medskip\medskip

In \secref{sec:reduction}, we relate higher-dimensional integrable scattering
to scattering in 1+1d integrable field theories.
We show that a particular one-dimensional reduction of the 2+1d integrable chiral model
leads to a mass deformation of the 1+1d Wess-Zumino-Witten model.
The deformation parameter in the 2d theory corresponds to
a choice of twisted boundary conditions along the compact direction in the 3d theory.
A certain subsector of the reduced theory is described by the 2d sine-Gordon model,
and the sine-Gordon lump solitons can be lifted to
a class of line solitons in the 2+1d integrable chiral model~\cite{Lee89}.
As a consistency check, we verify that our results for 3d line soliton scattering
reproduce the known result for 2d lump soliton scattering in the sine-Gordon model.

\section{3d Integrable Chiral Model}\label{sec:3dICM}
In this section, we will introduce the 3d integrable chiral model (3d ICM),
which was first studied by Manakov and Zakharov in~\cite{MZ81} and by Ward in~\cite{War88}.
These two papers studied different versions of the 3d ICM: the model depends on a choice of vector,
and this vector was taken to be timelike in~\cite{MZ81} and spacelike in~\cite{War88}.
This paper will focus on the case with a spacelike vector, following~\cite{War88}.

\medskip

Let us work on $\fR^{2,1}$ with coordinates $(t, x, y)$ in terms of which
the metric and volume form are written as $\dr s^2 = -\dr t^2 + \dr x^2 + \dr y^2$
and $\vol_3 = \dr t \wedge \dr x \wedge \dr y$.
The 3d integrable chiral model (3d ICM) is defined by the action
\begin{equation}
    S [g] =
    \frac{1}{2} \int_{\fR^3} \! \tr ( g^{-1} \dr g \wedge \star g^{-1} \dr g )
    + \frac{1}{3} \int_{\fR^3 \times [0,1]} \hspace{-2.2em} \dr y \wedge 
    \tr (\tilde{g}^{-1} \dr \tilde{g} \wedge \tilde{g}^{-1} \dr \tilde{g} \wedge \tilde{g}^{-1} \dr \tilde{g}) \,.
\end{equation}
The fundamental field $g : \fR^3 \to \grp{G}$ is a map from spacetime to a Lie group $\grp{G}$,
and $\tilde{g} : \fR^3 \times [0,1] \to \grp{G}$ denotes an extension of this map satisfying
$\tilde{g} \vert_1 = g$ and $\tilde{g} \vert_0 = g_0$ for some constant element $g_0 \in \grp{G}$.
The equation of motion (most succinctly expressed in terms of null coordinates
defined by $x = u + v$ and $t = u - v$) is given by
\begin{equation}
    \pd_u (\pd_v g g^{-1}) + \pd_y (\pd_y g g^{-1}) = 0 \,.
\end{equation}
Unlike the action, the equation of motion does not depend on the extension $\tilde{g}$,
showing that this choice does not affect the classical solutions of the theory.

\medskip

Spacetime Lorentz symmetry is partially broken by the explicit appearance of $\dr y$ in the action.
The subgroup of Lorentz transformations preserving this $1$-form is given by
$\grp{SO}(1,1)$ which acts as boosts in the $x$-direction.
Translations in spacetime are also symmetries of the theory, and the conserved energy is given by
\begin{equation}\label{eq:energy}
    E[g] = -\frac{1}{2} \int_{\fR^2} \dr x \wedge \dr y \,
    \tr \big( j_t^2 + j_x^2 + j_y^2 \big) \,, \qquad
    j_\mu = g^{-1} \pd_\mu g \,.
\end{equation}
The energy density is positive-definite for compact Lie groups,
and solutions of the theory can be visualised by plotting this density
over spatial slices for various fixed times.

\medskip

The theory possesses a $\grp{G} \times \grp{G}$ internal symmetry,
which acts on the field as
\begin{equation}
    g \mapsto h_\ell \, g \, h_r^{-1} \,.
\end{equation}
Each transformation parameter is allowed to depend on one of the null coordinates,
with semi-local symmetries parametrised by $h_\ell(v)$ and $h_r(u)$ respectively.
These transformations preserve the action, and map solutions of the theory to new solutions.

\medskip

Integrability can be understood in the Lax formalism, where the Lax pair is given by
\begin{equation}
    \nabla_1 = \pd_v - \pd_v g g^{-1} + z \, \pd_y \,, \qquad
    \nabla_2 = \pd_y - \pd_y g g^{-1} - z \, \pd_u \,.
\end{equation}
This pair of differential operators depends on an auxiliary complex variable $z \in \CP^1$,
which is often referred to as the spectral parameter.
The existence of solutions to the linear problem $\nabla U = 0$ is equivalent to the condition
$[\nabla_1, \nabla_2] = 0$, which implies that $g$ solves the equation of motion of the 3d ICM.

\medskip

Classical solutions of the 3d ICM are in 1-to-1 correspondence with solutions of
the 3d Bogomolny equation up to gauge transformations~\cite{War89}.
The Bogomolny equation for magnetic monopoles is given by $F = \star \Dr \phi$
where $F$ is the curvature of a connection $\Dr = \dr + A$ and
$\phi$ is a scalar field transforming in the adjoint representation.
In null coordinates on $\fR^{2,1}$, these equations are written as
\begin{equation}
    F_{uy} = - \Dr_u \phi \,, \qquad
    F_{uv} = 2 \, \Dr_y \phi \,, \qquad
    F_{vy} = \Dr_v \phi \,.
\end{equation}
These equations are integrable, and their Lax pair is given by
\begin{equation}
    \nabla_1 = \Dr_v + z \big( \Dr_y + \phi \big) \,, \qquad
    \nabla_2 = \Dr_y - \phi - z \, \Dr_u \,.
\end{equation}
To demonstrate the equivalence with the 3d ICM, note that the first equation can be rewritten as
$[\Dr_u, \Dr_y + \phi] = 0$ and the third equation can be rewritten as $[\Dr_v, \Dr_y - \phi] = 0$.
These equations imply the existence of fields $h$ and $\tilde {h}$ satisfying
\begin{equation}
    A_u = \tilde{h}^{-1} \pd_u \tilde{h} \,, \qquad
    A_y + \phi = \tilde{h}^{-1} \pd_y \tilde{h} \,, \qquad
    A_v = h^{-1} \pd_v h \,, \qquad
    A_y - \phi = h^{-1} \pd_y h \,.
\end{equation}
Performing a gauge transformation by $\tilde{h}$ takes us into a frame in which $A_u = 0$ and
$A_y + \phi = 0$, and the remaining fields take the form $A_v = -\pd_v g g^{-1}$ and
$A_y - \phi = -\pd_y g g^{-1}$ where $g = \tilde{h} h^{-1}$.
Substituting this configuration into the final Bogomolny equation
recovers the equation of motion of the 3d ICM.

\medskip

The 3d ICM can be derived as a one-dimensional reduction of
the self-dual Yang-Mills (SDYM) equation in $(2,2)$-signature~\cite{MZ81}.
Indeed, the Bogomolny equation itself is a one-dimensional reduction of the SDYM equation,
and the analogous equation to the 3d ICM in four dimensions is Yang's equation~\cite{Yan77},
which arises as the equation of motion of the 4d WZW model~\cite{Poh80,Don85,LMNS96}.
\begin{equation*}
\begin{tikzpicture}
\node at (0,0) {4d SDYM};
\draw[<->] (1.5,0) -- (3.75,0);
\node at (5,0) {4d WZW};
\draw[->] (0,-0.5) -- (0,-1);
\node at (0,-1.5) {3d Bogomolny};
\draw[<->] (1.5,-1.5) -- (3.75,-1.5);
\draw[->] (5,-0.5) -- (5,-1);
\node at (5,-1.5) {3d ICM};
\end{tikzpicture}
\end{equation*}
Taking a one-dimensional reduction of the 3d ICM along the $y$-direction gives the 2d WZW model,
while a reduction along the $x$-direction gives the Lorentzian principal chiral model (PCM).
Intermediate reductions along a generic direction in the $xy$-plane give
the PCM plus WZ-term with arbitrary couplings.
Reducing along the $t$-direction gives the Euclidean PCM,
whose equation of motion is the harmonic map equation.
Other integrable models, such as the 2d sine-Gordon model, can also be recovered as reductions
of the 3d ICM by choosing particular Lie groups $\grp{G}$ and imposing additional constraints.

\medskip

Solutions of the 3d ICM correspond to certain
holomorphic vector bundles over minitwistor space~\cite{Hit82,Hit83,War89},
and the finite energy solutions of the 3d ICM correspond to bundles
which extend over a compactification of minitwistor space~\cite{War90,Ana95}.
Minitwistor space $\MT$ is the total space of the bundle $\mathcal{O}(2) \to \CP^1$,
and we will denote a point by $(z, w) \in \MT$
where $z$ is the base coordinate and $w$ is the fibre coordinate.
The relationship with spacetime can be described
via the correspondence space $X = \CP^1 \times \fR^{2,1}$,
which comes equipped with a projection map $p : X \to \MT$ given by
\begin{equation}
    p : (z, u, v, y) \mapsto (z, w = u + z y - z^2 v) \,.
\end{equation}
A fixed point in $\fR^{2,1}$ defines a holomorphic embedding of $\CP^1$ into $\MT$,
and a fixed point in $\MT$ corresponds to an oriented timelike line in $\fR^{2,1}$.
This relationship between 3d spacetime and $\MT$ is known as
the minitwistor incidence relations~\cite{Hit82,Hit83,War89}.

\medskip

Furthermore, the 3d ICM can be constructed from 5d holomorphic-topological Chern-Simons (CS),
where the Lax of the 3d ICM is identified with the partial connection of 5d CS~\cite{BS23}.
The 5d CS setup relevant to the 3d ICM is defined by the action
\begin{equation}
    S_{\text{5dCS}} [a] = \frac{\iu}{2 \pi} \int_X \omega \wedge
    \tr \bigg( a \wedge \dr a + \frac{2}{3} a \wedge a \wedge a \bigg) \,, \qquad
    \omega = \frac{\dr z \wedge \dr w}{z^2} \,.
\end{equation}
The meromorphic $(2,0)$-form $\omega$ has a double pole at $z = 0$ and
a double pole at $\tilde{z} = 0$, where $\tilde{z} = z^{-1}$.
In order to see the double pole in the other patch, it is important to remember that
the fibre coordinate $w$ transforms as $\tilde{w} = z^{-2} w$
since it is valued in $\mathcal{O}(2)$.
It is necessary to impose boundary conditions of the gauge field at the poles of $\omega$,
and the boundary conditions relevant to the 3d ICM are given by
$a \vert_{z = 0} = 0$ and $a \vert_{\tilde{z} = 0} = 0$.

\section{Soliton Solutions}\label{sec:solitons}
In this section, we will study soliton solutions to the 3d ICM,
which can be constructed by solving a matrix Riemann-Hilbert problem with zeroes~\cite{MZ81,War88}.
This solution generating technique for integrable systems was introduced in~\cite{ZS74,ZS79},
and it is often referred to as the dressing method.
In the context of the 3d ICM with a spacelike vector, it can be used to construct lump solitons,
whose energy density is concentrated around a point in space
and whose total energy is finite~\cite{War88},
and line solitons, whose energy density is concentrated around a line in space
and whose total energy diverges~\cite{Lee89}.

\subsection{Solution Generating Technique}
Let us describe the solution generating technique known as the dressing method~\cite{ZS74,ZS79}.
Consider a partial connection (in the gauge $a_{\bar{z}} = 0$) which can be written as
\begin{equation}
    \nabla_{\bar{z}} = \pd_{\bar{z}} \,, \qquad
    \nabla_1 = \ell_1 + \Lax_1 \,, \qquad
    \nabla_2 = \ell_2 + \Lax_2 \,.
\end{equation}
For the 3d ICM, the $z$-dependent vector fields are
$\ell_1 = \pd_v + z \pd_y$ and $\ell_2 = \pd_y - z \pd_u$,
and the components of the connection are
$\Lax_1 = - \pd_v g g^{-1}$ and $\Lax_2 = - \pd_y g g^{-1}$.
The vector fields are constructed to satisfy $\ell_I (w_z) = 0$ for $I \in \{ 1, 2 \}$,
where $w_z = u + z y - z^2 v$ is the holomorphic fibre coordinate on minitwistor space.
In fact, this presentation of the Lax applies to a large selection of integrable systems
across various dimensions, with different choices of the vector fields $\ell_I$ and different
expressions for the connection components $\Lax_I$ corresponding to different integrable models.

\medskip

Gauge transformations $H : \CP^1 \times \fR^{2,1} \to \grp{G}$
act on the partial connection $\nabla$
and the solution $U$ to the linear problem $\nabla U = 0$ as
\begin{equation}
    \nabla \mapsto H^{-1} \nabla H \,, \qquad
    U \mapsto H^{-1} U \,.
\end{equation}
These transformations preserve the condition $[\nabla_1, \nabla_2] = 0$, and therefore
stand a chance of generating new solutions to the associated integrable system.
In order to preserve the gauge fixing condition $a_{\bar{z}} = 0$, the gauge transformation
must satisfy $H^{-1} \pd_{\bar{z}} H = 0$.
The simplest solutions to this condition are found by taking $H$ to be holomorphic,
in which case Liouville's theorem states that $H$ must be constant in the $z$-plane.

\medskip

Dressing transformations are found by allowing $H$ to be meromorphic in the $z$-plane,
at which point the problem amounts to constructing meromorphic matrices $H$ and $H^{-1}$
satisfying $H^{-1} H = \id$.
In other words, the problem is to factorise the identity matrix into two meromorphic matrices
$H$ and $H^{-1}$ on $\CP^1$ which is known as a matrix Riemann-Hilbert problem with zeroes.

\medskip

Let us assume that $H$ has simple poles at $\alpha_i \in \CP^1$ for $i \in \{1, \dots, k \}$,
and that $H^{-1}$ has simple poles at $\beta_j \in \CP^1$ for $j \in \{1, \dots, k \}$,
where all of these points are distinct.
For the moment, $H^{-1}$ should be thought of as notation for an arbitrary matrix,
and only later will the constraint $H^{-1} H = \id$ be imposed.
These matrices can be written as
\begin{equation}
    H = \id + \sum_{i=1}^k \frac{A_i}{z - \alpha_i} \,, \qquad
    H^{-1} = \id + \sum_{j=1}^k \frac{B_j}{z - \beta_j} \,.
\end{equation}
These are the most general expressions for meromorphic matrices with the prescribed poles,
except that the additional condition $H \vert_{z = \infty} = \id$ has been imposed.
This turns out to be appropriate for the 3d ICM,
but it may be convenient to relax this condition for other integrable systems.

\medskip

Solutions to the constraint $H^{-1} H = \id$ can be found by assuming that the residues
$A_i$ and $B_j$ are degenerate matrices of rank $r_i \leq N$ and $r_j \leq N$,
where the gauge group $\grp{G}$ is a subset of $\grp{GL}(N, \fC)$.
The rank decomposition of these matrices can be written as
\begin{equation}
    A_i = a_i \, X_i \,, \qquad
    B_j = Y_j \, b_j \,.
\end{equation}
For example, $a_i$ is an $(N \times r_i)$-matrix and $X_i$ is an $(r_i \times N)$-matrix.
At this stage, it is possible to solve the condition $H^{-1} H = \id$ for
$X_i$ and $Y_j$ in terms of $a_i$ and $b_j$, thereby solving the matrix Riemann-Hilbert problem.
For simplicity, let us present the solution when all of the residues are rank $1$ matrices.

\medskip

The combination $H^{-1} H$ is a meromorphic matrix on $\CP^1$ which coincides with
the identity matrix at infinity.
If this combination is in fact holomorphic, then Liouville's theorem states that it is constant
on $\CP^1$, and therefore solves the condition $H^{-1} H = \id$.
This is equivalent to the condition that all of the residues vanish,
which amounts to the following system of equations.
\begin{equation}\begin{aligned}
    \res_{\alpha_i} (H^{-1} H) & = 0 \quad & & \iff & \quad
    a_i + \sum_{j=1}^k \frac{b_j \cdot a_i}{\alpha_i - \beta_j} \, Y_j & = 0 \,, \\
    \res_{\beta_j} (H^{-1} H) & = 0 \quad & & \iff & \quad
    b_j - \sum_{i=1}^k \frac{b_j \cdot a_i}{\alpha_i - \beta_j} \, X_i & = 0 \,.
\end{aligned}\end{equation}
These equations can be solved for $X_i$ and $Y_j$ in terms of $a_i$ and $b_j$ by
inverting the $(k \times k)$-matrix with entries
$M_{ij} = \frac{b_j \cdot a_i}{\alpha_i - \beta_j}$.
This completes the solution to the matrix Riemann-Hilbert problem with zeroes.

\medskip

It is useful to consider the case $k = 1$, where the solutions are given by
\begin{equation}
    H = \id - \frac{\beta - \alpha}{z - \alpha} \, P \,, \qquad
    H^{-1} = \id - \frac{\alpha - \beta}{z - \beta} \, P \,, \qquad
    P = \frac{a \, b}{b \cdot a} \,.
\end{equation}
The rank-$1$ projector $P$ satisfies $P^2 = P$, and the complementary projector
$P^\perp = \id - P$ satisfies $P P^\perp = 0$ and $P^\perp P = 0$.
The combination $H^{-1} H$ does not have any poles because $H^{-1}$ is
equal to $P^\perp$ at the pole of $H$ and vice versa.
In addition, the matrices $H$ and $H^{-1}$ only depend on the $N$-vectors $a$ and $b$
through the projector $P$, which is invariant under independent rescalings of these vectors.
In other words, the solution only depends on the $N$-vectors $a$ and $b$ up to rescalings.

\medskip

This simple case is especially useful to keep in mind because solutions with $k > 1$
can be rewritten as the composition of elementary $k = 1$ solutions.
Indeed, the general solution can be presented as
\begin{equation}
    H = \prod_{i=1}^k \bigg( \id - \frac{\beta_i - \alpha_i}{z - \alpha_i} \, P_i \bigg) \,, \qquad
    H^{-1} = \prod_{i=1}^k \bigg( \id - \frac{\alpha_i - \beta_i}{z - \beta_i} \, P_i \bigg) \,.
\end{equation}
It is important to choose a consistent ordering for the matrix products in these expressions,
and the projectors $P_i$ will depend on this choice of ordering.
Each of these presentations of the general solution can be useful in different circumstances.

\medskip

Having found some solutions to the matrix Riemann-Hilbert problem,
let us return to the interpretation of these transformations
as symmetries of the integrable system.
The transformed Lax connection $H^{-1} \nabla H$ will still satisfy the condition
$[\nabla_1 , \nabla_2 ] = 0$ because this is preserved by gauge transformations.
However, this does not guarantee that the transformed Lax will remain a valid Lax
for the particular integrable system in question.
The Lax connections of different integrable systems have different analytic properties,
such as the location of their poles in the $z$-plane.
In order to preserve this feature, ensuring that the transformed Lax remains a valid Lax
for the same integrable system, the dressing transformation must not introduce any new poles.
In other words, the residues of the transformed Lax must vanish at the poles of $H$ and $H^{-1}$.
\begin{equation}\begin{aligned}
    \res_{\alpha_i} (H^{-1} \nabla_I H) & = 0 \quad & & \iff & \quad
    \ell_I \vert_{\alpha_i} (a_i) + (\Lax_I \vert_{\alpha_i}) \, a_i & = 0 \,, \\
    \res_{\beta_j} (H^{-1} \nabla_I H) & = 0 \quad & & \iff & \quad
    \ell_I \vert_{\beta_j} (b_j) - b_j \, (\Lax_I \vert_{\beta_j}) & = 0 \,.
\end{aligned}\end{equation}
Solving these equations requires knowing the solution to the linear problem $\nabla U = 0$.
Given this solution, the $N$-vectors $a_i$ and $b_j$ can be written as
\begin{equation}\begin{aligned}
    a_i & = (U \vert_{\alpha_i}) \, \hat{a}_i \,, \quad & \quad
    \ell_I \vert_{\alpha_i} (\hat{a}_i) & = 0 \,, \\
    b_j & =  \hat{b}_j \, (U^{-1} \vert_{\beta_j}) \,, \quad & \quad
    \ell_I \vert_{\beta_j} (\hat{b}_j) & = 0 \,.
\end{aligned}\end{equation}
The $N$-vectors $\hat{a}_i$ and $\hat{b}_j$ are undetermined parameters
which reflect the freedom in solving $\nabla U = 0$.
Notably, the parameters of the gauge transformation $H$ depend on the 3d field
through the solution $U$ to the linear problem.
In this sense, these gauge transformations are field-dependent.

\medskip

Lax connections of specific integrable systems are also subject to reality conditions,
which must be preserved by dressing transformations.
\unskip\footnote{
Reality conditions for Lax connections of integrable systems
are discussed, for example, in~\cite{MW96}.
}
These reality conditions are constructed from two pieces of data:
an antiholomorphic involution $C : \CP^1 \to \CP^1$ of the $z$-plane,
and a conjugate-linear automorphism $\Theta : \grp{G} \to \grp{G}$ of the gauge group.
The automorphism $\Theta$ specifies a real form of the gauge group;
for example, unitary matrices are the subgroup of $\grp{GL}(N,\fC)$ that are invariant
under $\Theta : U \mapsto (U^{-1})^\dagger$.
Meanwhile, the involution $C$ is associated with the signature of spacetime via twistor theory.
For example, split signature spacetime $\fR^{2,2}$ (which reduces to $\fR^{2,1}$) is specified by
the involution $C : z \mapsto \bar{z}$.
The reality condition on the Lax connection is then written as $C^\ast (\Lax) = \Theta (\Lax)$,
which is preserved by dressing transformations satisfying $C^\ast (H) = \Theta (H)$
or equivalently $\beta_i = \bar\alpha_i$ and $b_i = a_i^\dagger$.

\medskip

In summary, dressing transformations are singular field-dependent gauge transformations
which are built from a solution to a matrix Riemann-Hilbert problem with zeroes.
The parameters of this transformation are the locations of $k$ simple poles
$\alpha_i \in \CP^1$ in the $z$-plane,
together with an $N$-vector $\hat{a}_i$ for each of these poles
which is constrained to satisfy $\ell_I \vert_{\alpha_i} (\hat{a}_i) = 0$.
The elementary case ($k = 1$) takes the form
\begin{equation}
    H = \id - \frac{\bar\alpha - \alpha}{z - \alpha} \, P \,, \qquad
    H^{-1} = \id - \frac{\alpha - \bar\alpha}{z - \bar\alpha} \, P \,, \qquad
    P = \frac{a \, a^\dagger}{\Vert a \Vert^2} \,,
\end{equation}
where $a = (U \vert_{\alpha}) \hat{a}$ is built from the solution $U$
to the linear problem $\nabla U = 0$.
The role of the $N$-vector is to specify the Hermitian rank-$1$ projector $P$,
which is equivalent to an element of the complex Grassmannian $\cat{Gr}(1,N)$.
More generally, a Hermitian rank-$r$ projector $P$ is equivalent to an element of $\cat{Gr}(r,N)$.
These transformations provide a solution generating technique in the field of integrable systems.

\subsection{General Properties}
Let us apply this solution generating technique to the 3d integrable chiral model and
explore the space of solutions of this theory.
The input data of a dressing transformation includes an existing solution of the theory,
together with its Lax connection and the solution to the linear problem $\nabla U = 0$.
Let us start by applying a dressing transformation to the trivial solution $g = \id$
whose Lax is simply $\Lax_I = 0$, implying that $U = \id$ solves the linear problem.

\medskip

Working with the reality conditions for $\fR^{2,1}$ and the subgroup of unitary matrices,
a single dressing transformation generates the new solution
\begin{equation}
    U = \id - \frac{\alpha - \bar\alpha}{z - \bar\alpha} \, P \,, \qquad
    P = \frac{a \, a^\dagger}{\Vert a \Vert^2} \,.
\end{equation}
Comparing the new Lax connection $\Lax_I = - \ell_I U U^{-1}$ with its expression in terms of the
3d field $g$ shows that the new solution to the 3d ICM is given by evaluating $U$ at $z = 0$.
Adopting polar coordinates $\alpha = r e^{\iu \theta}$ for the pole of the dressing transformation,
the 3d solution can be written as
\begin{equation}
    g = P^\perp + e^{2 \iu \theta} P \,, \qquad
    P^\perp = \id - P \,.
\end{equation}
Let us highlight two features of this solution.
First, this solution is a map from $\fR^{2,1}$ to the group of unitary matrices $\grp{U}(N)$.
However, in all of the examples we will consider, the determinant is constant in spacetime,
so it is possible to rescale $g$ and produce a solution valued in $\grp{SU}(N)$.
Second, all of the spacetime-dependence is captured by the $N$-vector $a$, which is
a holomorphic function of the minitwistor coordinate $w_\alpha = u + \alpha y - \alpha^2 v$.
Taking into account the invariance of the solution under rescalings of $a$,
the physical data is actually a holomorphic map from the $w$-plane to $\CP^{N-1}$.
Let us specialise to the target space $\grp{U}(2)$ for simplicity.

\medskip

When $N = 2$, the rescaling symmetry can be used to set $a = (1, f)$,
where $f(w_\alpha)$ is a holomorphic function
of the minitwistor coordinate $w_\alpha = u + \alpha y - \alpha^2 v$,
leading to the explicit form of the solution
\begin{equation}
    g = \frac{e^{\iu \theta}}{1 + \vert f \vert^2}
    \begin{pmatrix}
        e^{\iu \theta} + e^{-\iu \theta} \vert f \vert^2    & 2 \, \iu \sin(\theta) \bar{f} \\
        2 \, \iu \sin(\theta) f                             & e^{-\iu \theta} + e^{\iu \theta} \vert f \vert^2
    \end{pmatrix} \,.
\end{equation}
Some features of this solution can be understood by considering the following set of coordinates
on the target space $\grp{U}(2) \cong S^1 \times S^3$.
\begin{equation}\begin{aligned}
    &
    \begin{aligned}
        X_1 & = \cos(\phi_1) \\
        X_2 & = \sin(\phi_1) \, \frac{1 - \vert Z \vert^2}{1 + \vert Z \vert^2}
    \end{aligned}
    \quad & \quad &
    U = e^{\iu \phi_0}
    \begin{pmatrix}
        X_1 + \iu X_2   & X_3 + \iu X_4 \\
        -X_3 + \iu X_4  & X_1 - \iu X_2
    \end{pmatrix}
    \\
    & X_3 = \sin(\phi_1) \, \frac{-\iu (Z - \bar{Z})}{1 + \vert Z \vert^2}
    \quad & \quad &
    \qquad X_1^2 + X_2^2 + X_3^2 + X_4^2 = 1 \\
    & X_4 = \sin(\phi_1) \, \frac{Z + \bar{Z}}{1 + \vert Z \vert^2}
    \quad & \quad &
    \dr s^2 = \dr \phi_0^2 + \dr \phi_1^2 + 4 \sin^2 (\phi_1) \frac{\dr Z \dr \bar{Z}}{(1 + \vert Z \vert^2)^2}
\end{aligned}\end{equation}
In this expression, $U$ is a general element of $\grp{U}(2)$ which is parametrised by
the target space coordinates $( \phi_0, \phi_1, Z, \bar{Z})$.
The determinant factor is captured by $\phi_0$ (which ranges from $0$ to $2 \pi$),
and the remaining coordinates parametrise $\grp{SU}(2) \cong S^3$.
For any fixed value of $\phi_1$ (between $0$ and $\pi$),
the complex coordinate $Z$ parametrises a $\CP^1 \cong S^2$ of radius $\sin (\phi_1)$.
The metric $\dr s^2 = -\frac{1}{2} \tr(U^{-1} \dr U U^{-1} \dr U)$ is also given explicitly above.
In terms of these coordinates, the 3d solution is written as
\begin{equation}
    \phi_0 = \theta \,, \qquad
    \phi_1 = \theta \,, \qquad
    Z = f(w) \,.
\end{equation}
The 3d solution $g : \fR^{2,1} \to \grp{U}(2)$ does not explore all of $\grp{U}(2)$,
as its image is contained within the $\CP^1 \subset \grp{U}(2)$, defined by
$\{ \phi_0 = \theta , \phi_1 = \theta \} \subset \grp{U}(2)$ and parametrised by $Z$.

\medskip

The total conserved energy~\eqref{eq:energy} of this solution is given by
\begin{equation}
    E[g] = 4 \sin^2 (\theta) \int_{\fR^2} \dr x \wedge \dr y \,
    \frac{\delta^{\mu \nu} \pd_\mu f \pd_\nu \bar{f}}{(1 + \vert f \vert^2)^2}
    = 4 \, \frac{1 + r^2}{r} \vert \sin(\theta) \vert \int_{\fC} f^\ast (\vol_{\CP^1}) \,.
\end{equation}
For a fixed choice of $\alpha = r e^{\iu \theta}$, the total energy is quantised by the degree
of $f : \fC \to \CP^1$ which counts the number of times that the image $f(\fC)$ wraps $\CP^1$.
The volume of $\CP^1$ is $\pi$, so the integral in the second expression gives precisely
$\int_{\fC} f^\ast (\vol_{\CP^1}) = \pi \deg (f)$.
When this degree is finite, it counts the number of lumps in the energy density for generic maps.
For example, let us consider the simplest nontrivial map $f(w) = w$,
whose energy density is proportional to
\begin{equation}
    \frac{\pd_w f \pd_{\bar{w}} \bar{f}}{(1 + \vert f \vert^2)^2} =
    \frac{1}{(1 + \vert w \vert^2)^2} \,.
\end{equation}
This energy density is peaked at $w = 0$, and tends to zero as $w \to \infty$.
Since the energy density is localised around a point,
this solution can be interpreted as a single lump soliton centred at $w = 0$.

\medskip

The trajectory of this lump soliton in spacetime is defined by the equation $w = 0$,
which implies that its velocity is given by
\begin{equation}
    v_x = \frac{-1 + r^2}{1 + r^2} \,, \qquad
    v_y = -\frac{2 \, r \cos(\theta)}{1 + r^2} \,, \qquad
    \Vert v \Vert^2 = 1 - \frac{4 \, r^2 \sin^2(\theta)}{(1 + r^2)^2} \,.
\end{equation}
This velocity is always less than the speed of light ($c = 1$), and for fixed $\theta$
it varies between $\cos^2 (\theta) \leq \Vert v \Vert^2 < 1$ with the minimum at $r = 1$.
While the profile of the soliton in the $w$-plane is determined by the map $f : \fC \to \CP^1$, its
movement in spacetime is determined by the pole of the dressing transformation $\alpha \in \CP^1$.
As shown in~\cite{War88}, the total energy of the soliton can be written as
$E = 8 \pi \gamma \sin^2 (\theta) \deg(f)$ in terms of
the Lorentz factor $\gamma = 1 / \sqrt{1 - \Vert v \Vert^2}$.
This relationship is relativistic when $\theta = \frac{\pi}{2}$ and
the soliton moves exclusively along the $x$-axis,
but it departs from the relativistic case when
the soliton acquires a velocity in the $y$-direction~\cite{War88}.

\medskip

Solutions of the linear problem $\nabla U = 0$ determine
holomorphic vector bundles over minitwistor space~\cite{Hit82,Hit83,War89},
and the finite energy solutions correspond to bundles
which extend over a compactification of minitwistor space~\cite{War90,Ana95}.
Minitwistor space is a two-dimensional complex manifold, and holomorphic vector bundles over
its compactification can have a nontrivial second Chern number.
This topological invariant is quantised, and it was shown to be proportional to
the quantised energy for time-independent solutions in~\cite{Ana95}.
For time-dependent solutions, the energy was shown to be proportional to
the third homotopy class of (a restriction of) $U$ in~\cite{DP07},
and this was matched with the second Chern number of the associated bundles in~\cite{Pla16}.

\subsection{Lump Solitons}
\begin{figure}
\begin{subfigure}{0.45\textwidth}
\centering
\includegraphics[width=\textwidth]{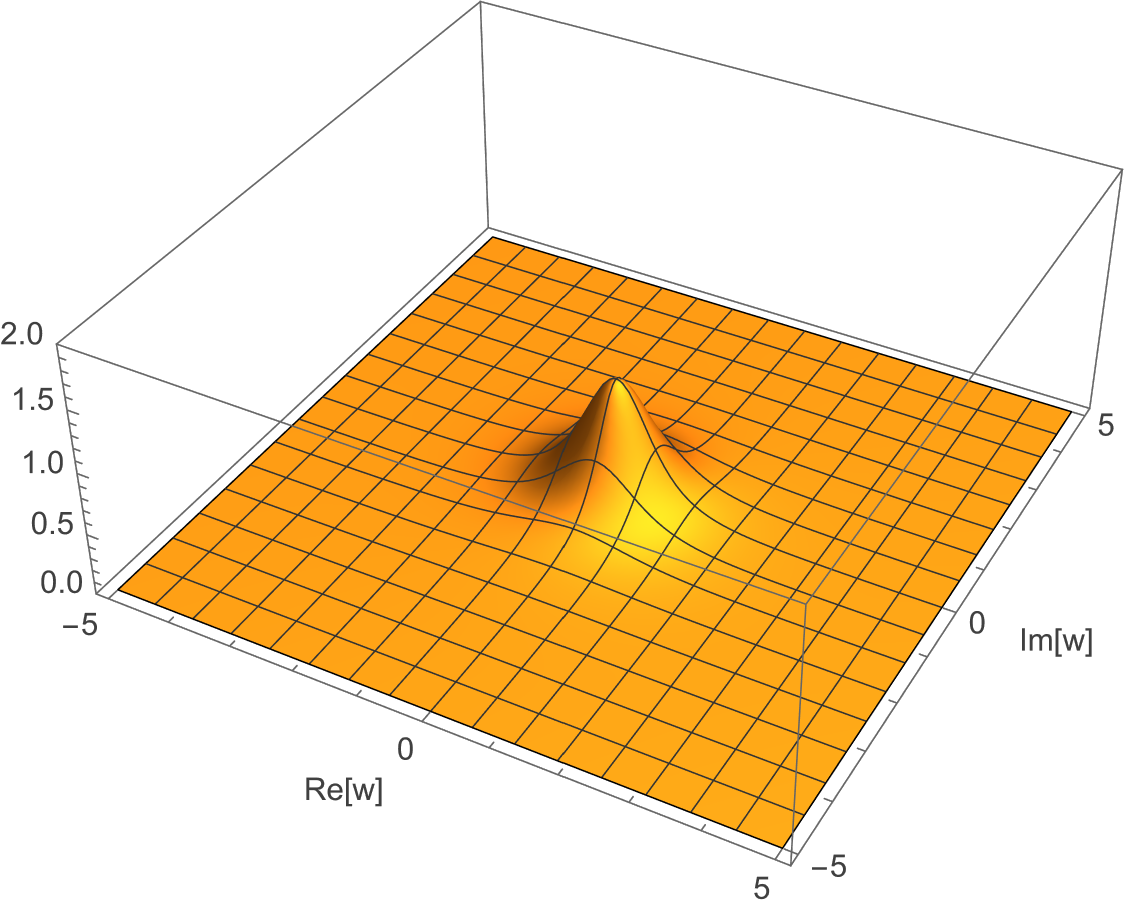}
\caption{$f(w) = w$.}
\end{subfigure}%
\hfill
\begin{subfigure}{0.45\textwidth}
\centering
\includegraphics[width=\textwidth]{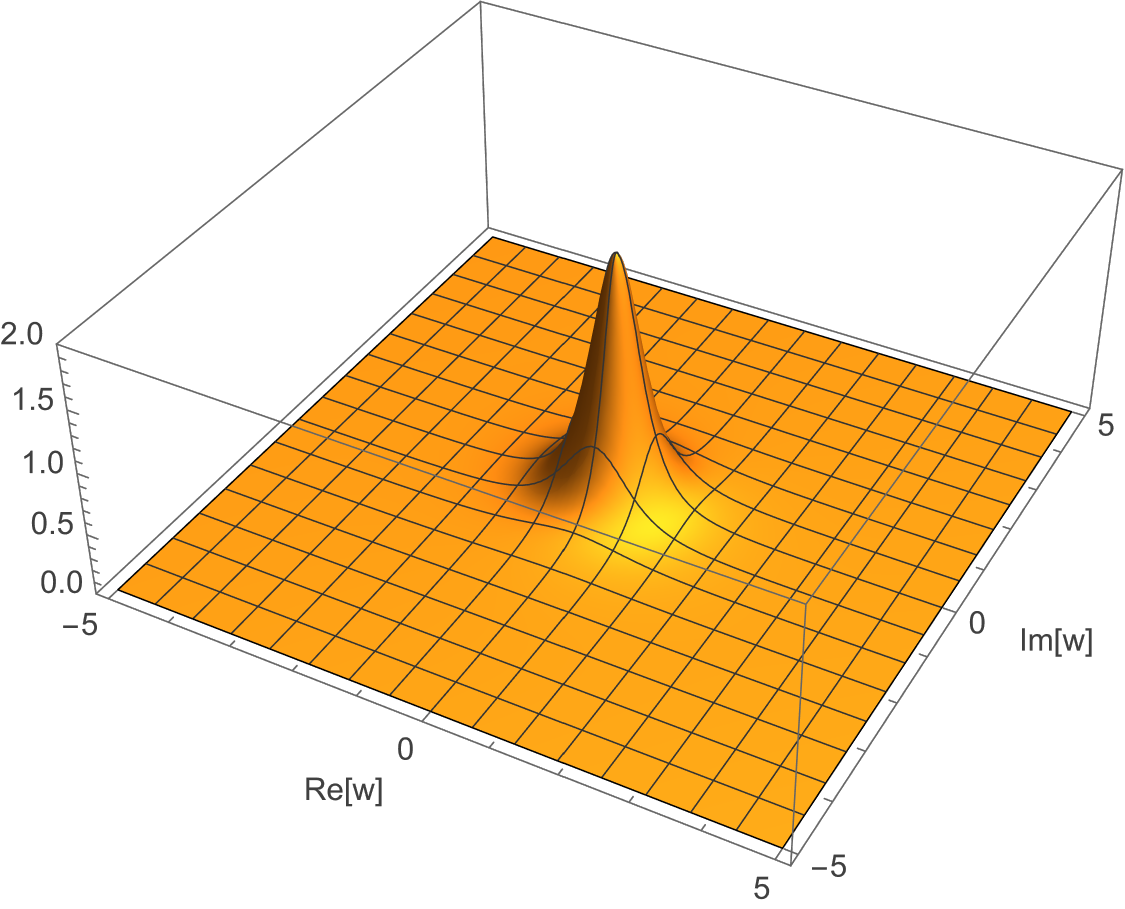}
\caption{$f(w) = \sqrt{2} \, w$.}
\end{subfigure}%
\vspace{1.5em}\\%
\begin{subfigure}{0.45\textwidth}
\centering
\includegraphics[width=\textwidth]{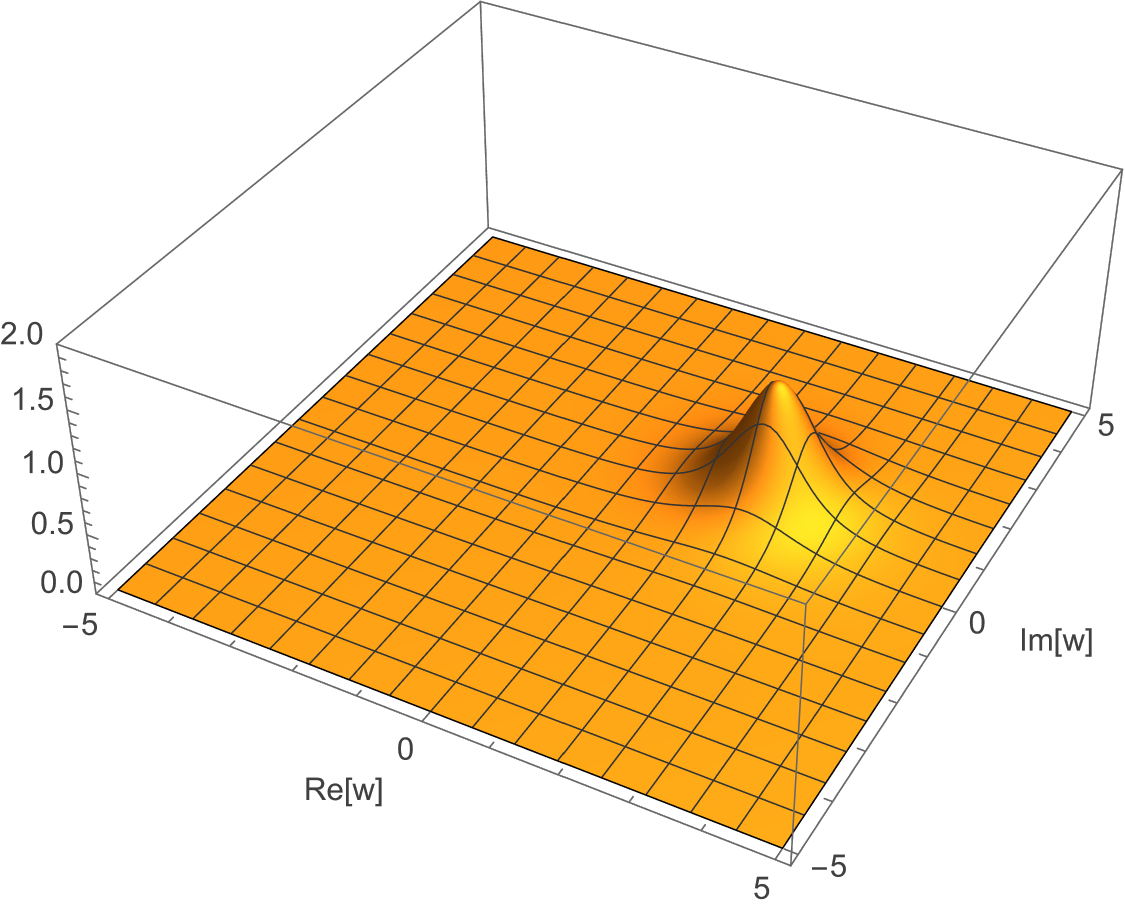}
\caption{$f(w) = w - (2 + \iu)$.}
\end{subfigure}%
\hfill
\begin{subfigure}{0.45\textwidth}
\centering
\includegraphics[width=\textwidth]{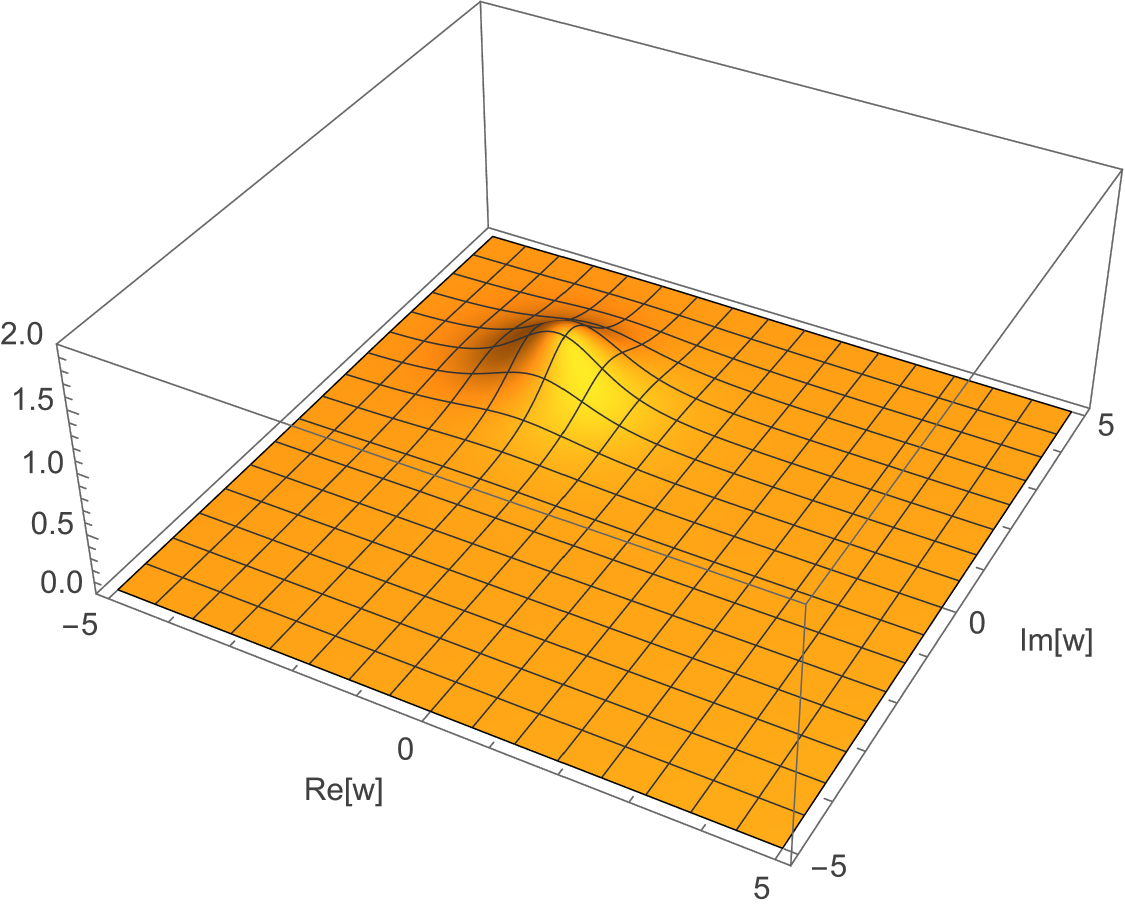}
\caption{$f(w) = \frac{1}{\sqrt{2}} \big(w - (-2 + 2 \iu) \big)$.}
\end{subfigure}%
\vspace{1em}
\caption{
    Degree-1 maps of the form $f(w) = c \, (w - \protect\rchi_0)$ correspond to 1-lump solitons
    centred at $w = \protect\rchi_0$ with a height proportional to $\vert c \vert^2$.
    The energy density is proportional to $\frac{\vert f^\prime \vert^2}{(1 + \vert f \vert^2)^2}$,
    which has been plotted over the $w$-plane for various examples.
}
\label{fig:1lumps}
\end{figure}

Let us consider the moduli space of the lump soliton solutions,
meaning the moduli space of rational holomorphic maps from $\fC$ to $\CP^1$.
The degree of a rational function $f(w) = p(w) / q(w)$ is defined to be
the maximum degree of the two polynomials $p(w)$ and $q(w)$.
The most general degree-$1$ function can be written as
\begin{equation}
    f(w) = c \, \frac{w - \rchi_0}{w - \rchi_\infty} \,.
\end{equation}
However, the energy density is not sensitive to all of these parameters as it is invariant
under the global $\grp{U}(2) \times \grp{U}(2)$-action.
The diagonal $\grp{SU}(2)$ subgroup acts on the target space $\CP^1$ as rigid rotations,
and this can be used to fix three real parameters in $f$.
Let us put the pole of $f(w)$ at $w = \infty$ and fix $c \in \fR$, giving the degree-$1$ map
\begin{equation}
    f(w) = c \, (w - \rchi_0) \,.
\end{equation}
The energy density of this solution is proportional to
\begin{equation}
    \frac{\pd_w f \pd_{\bar{w}} \bar{f}}{(1 + \vert f \vert^2)^2} =
    \frac{\vert c \vert^2}{(1 + \vert c \vert^2 \vert w - \rchi_0 \vert^2)^2} \,.
\end{equation}
The remaining parameters in $f$ have a direct physical interpretation:
the lump soliton is centred at $w = \rchi_0$ and the height is proportional to $\vert c \vert^2$.
The width of the soliton shrinks as its height increases, keeping the total energy fixed.
Some examples of degree-$1$ maps are presented in~\figref{fig:1lumps}.

\begin{figure}
\begin{subfigure}{0.45\textwidth}
\centering
\includegraphics[width=\textwidth]{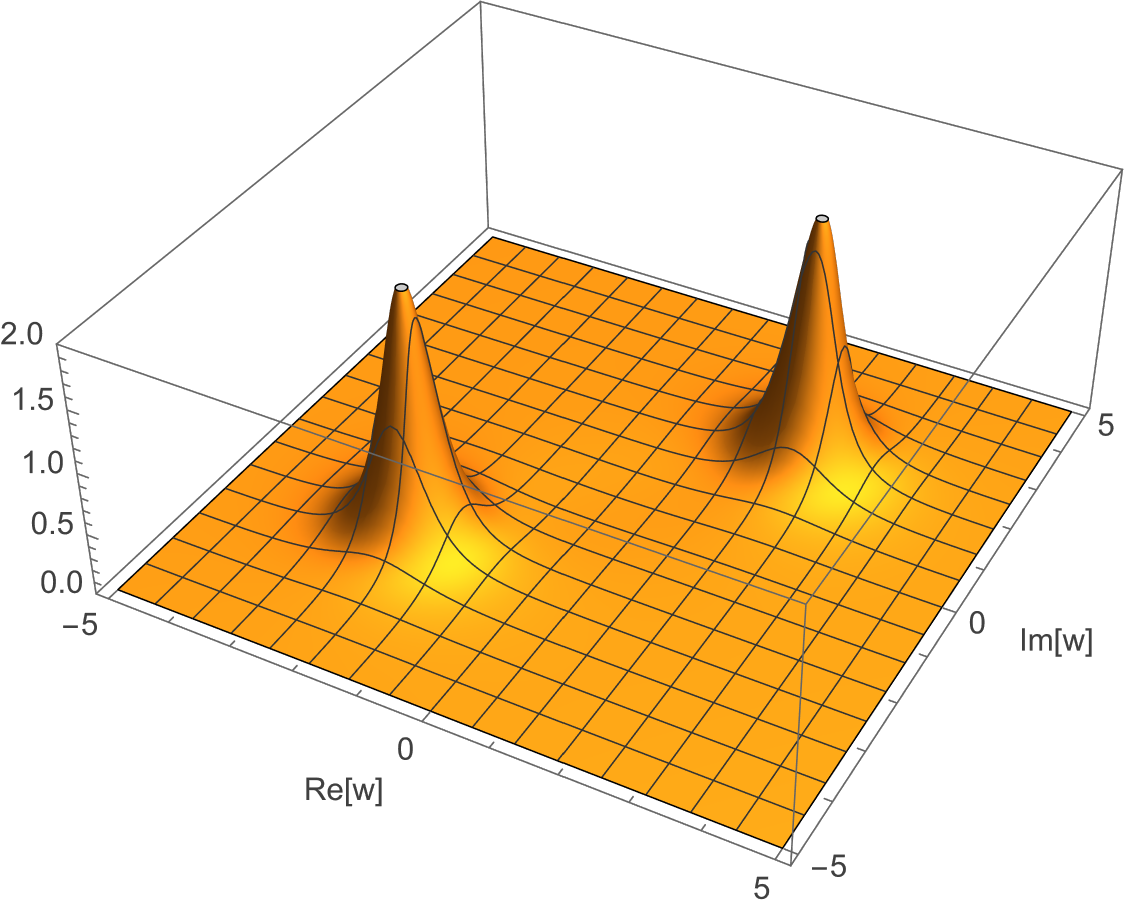}
\caption{$f(w) = \frac{1}{4} \big( w^2 - (2 + 2 \iu)^2 \big)$.}
\end{subfigure}%
\hfill
\begin{subfigure}{0.45\textwidth}
\centering
\includegraphics[width=\textwidth]{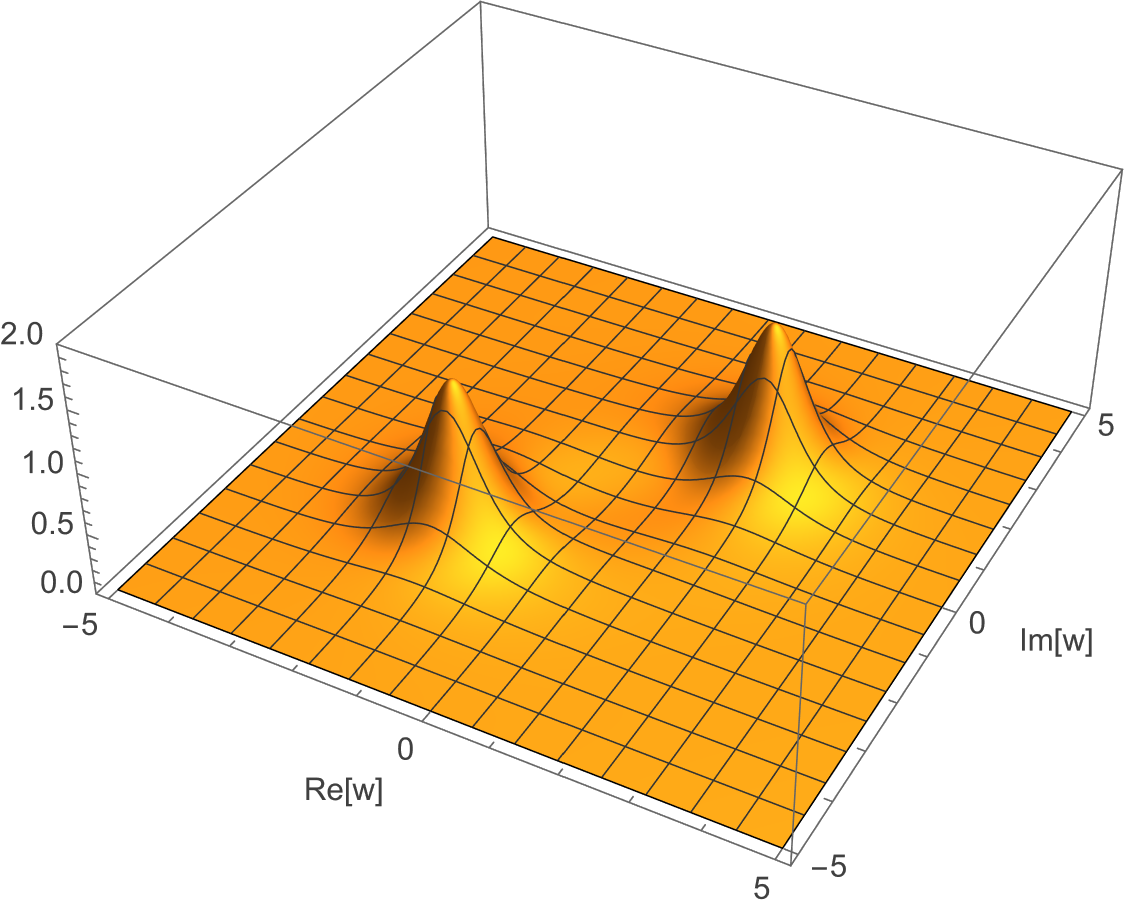}
\caption{$f(w) = \frac{1}{4} \big( w^2 - (\tfrac{3}{2} + \tfrac{3}{2} \iu)^2 \big)$.}
\end{subfigure}%
\vspace{1.5em}\\%
\begin{subfigure}{0.45\textwidth}
\centering
\includegraphics[width=\textwidth]{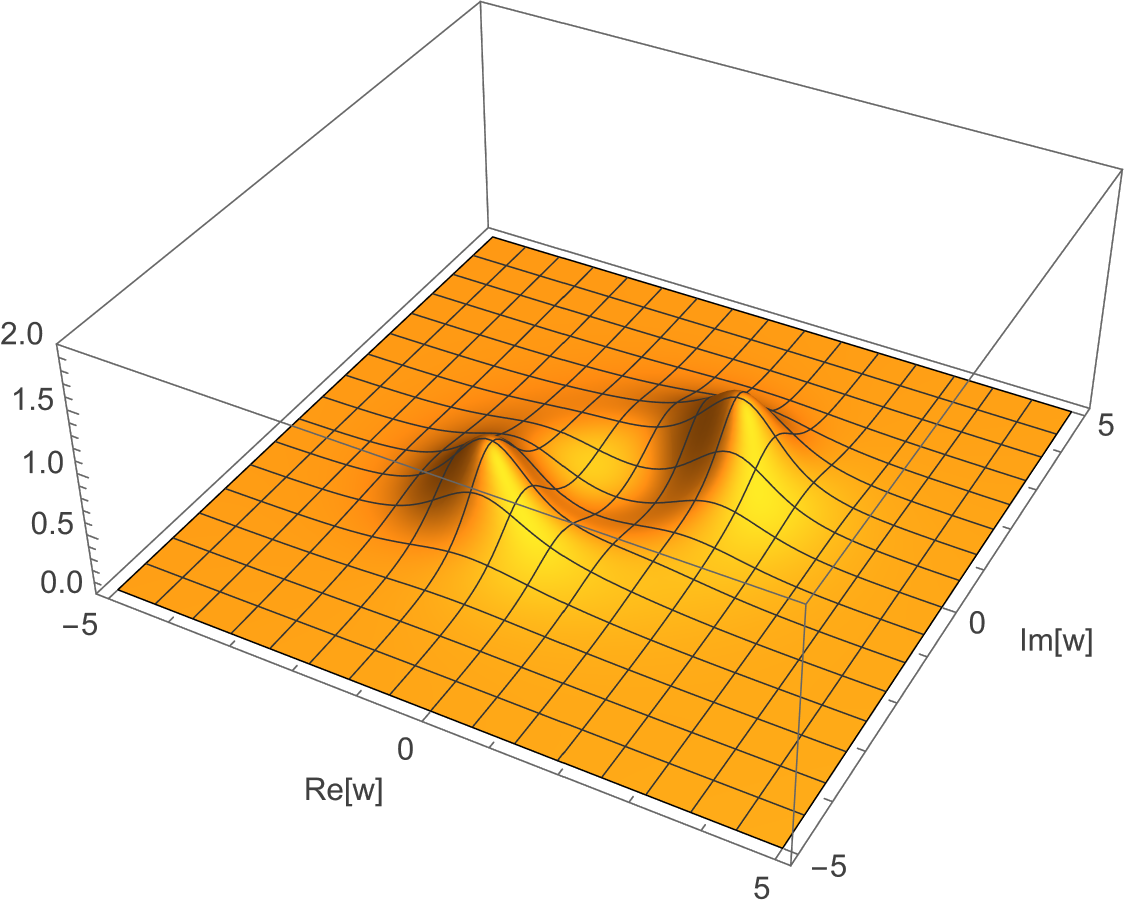}
\caption{$f(w) = \frac{1}{4} \big( w^2 - (1 + \iu)^2 \big)$.}
\end{subfigure}%
\hfill
\begin{subfigure}{0.45\textwidth}
\centering
\includegraphics[width=\textwidth]{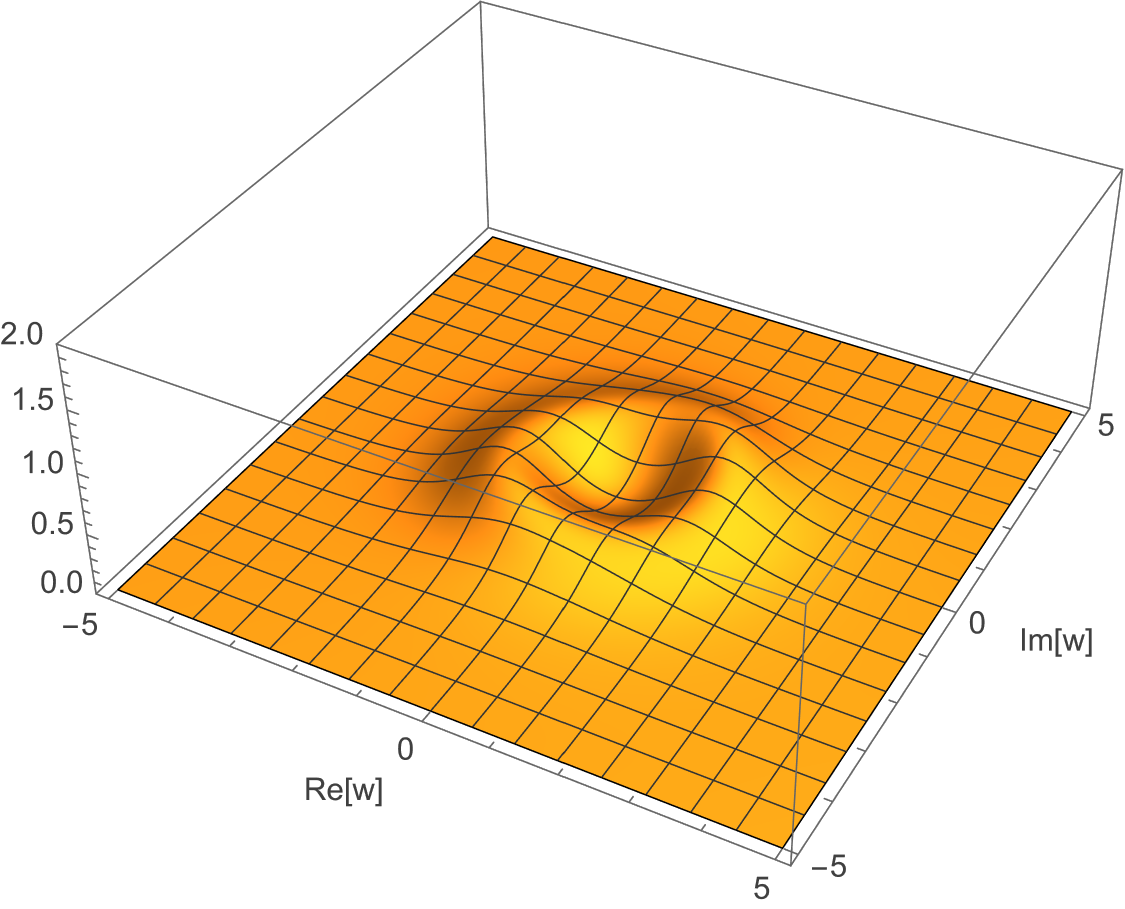}
\caption{$f(w) = \frac{1}{4} \big( w^2 - (\tfrac{1}{2} + \tfrac{1}{2} \iu)^2 \big)$.}
\end{subfigure}%
\vspace{1em}
\caption{
    Generically, degree-$2$ maps of the form
    $f(w) = c \, (w^2 - \protect\rchi_\Delta^2)$ correspond to
    2-lump solitons centred at $w = \pm \protect\rchi_\Delta$.
    However, as the two lumps approach each other they form a ring-shaped soliton.
    The energy density is proportional to $\frac{\vert f^\prime \vert^2}{(1 + \vert f \vert^2)^2}$,
    which has been plotted over the $w$-plane for various examples.
}
\label{fig:2lumps}
\end{figure}

\medskip

A general degree-$n$ rational function has $4 n + 2$ real parameters,
which leaves $4 n - 1$ parameters after taking into account the $\grp{SU}(2)$-invariance.
Generically, these maps describe $n$-lump soliton solutions and $2 n$ of the parameters encode
the locations of these lumps in the $w$-plane.
However, when two lumps approach each other, they form a ring-shaped soliton
in which the distinction between the two lumps is lost.
This behaviour is presented for some degree-$2$ maps in~\figref{fig:2lumps}.

\subsection{Line Solitons}
\begin{figure}
\begin{subfigure}{0.45\textwidth}
\centering
\includegraphics[width=\textwidth]{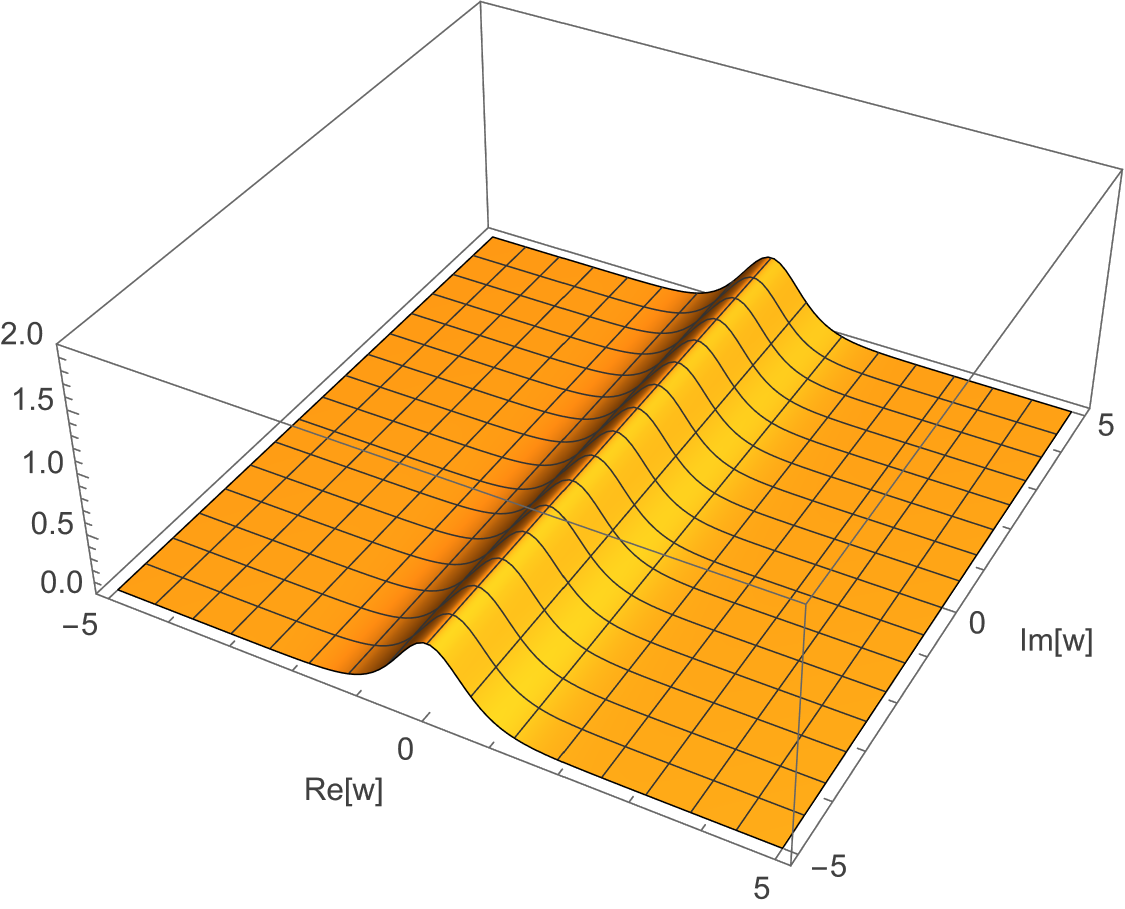}
\caption{$f(w) = \exp(\frac{3}{2} w)$.}
\end{subfigure}%
\hfill
\begin{subfigure}{0.45\textwidth}
\centering
\includegraphics[width=\textwidth]{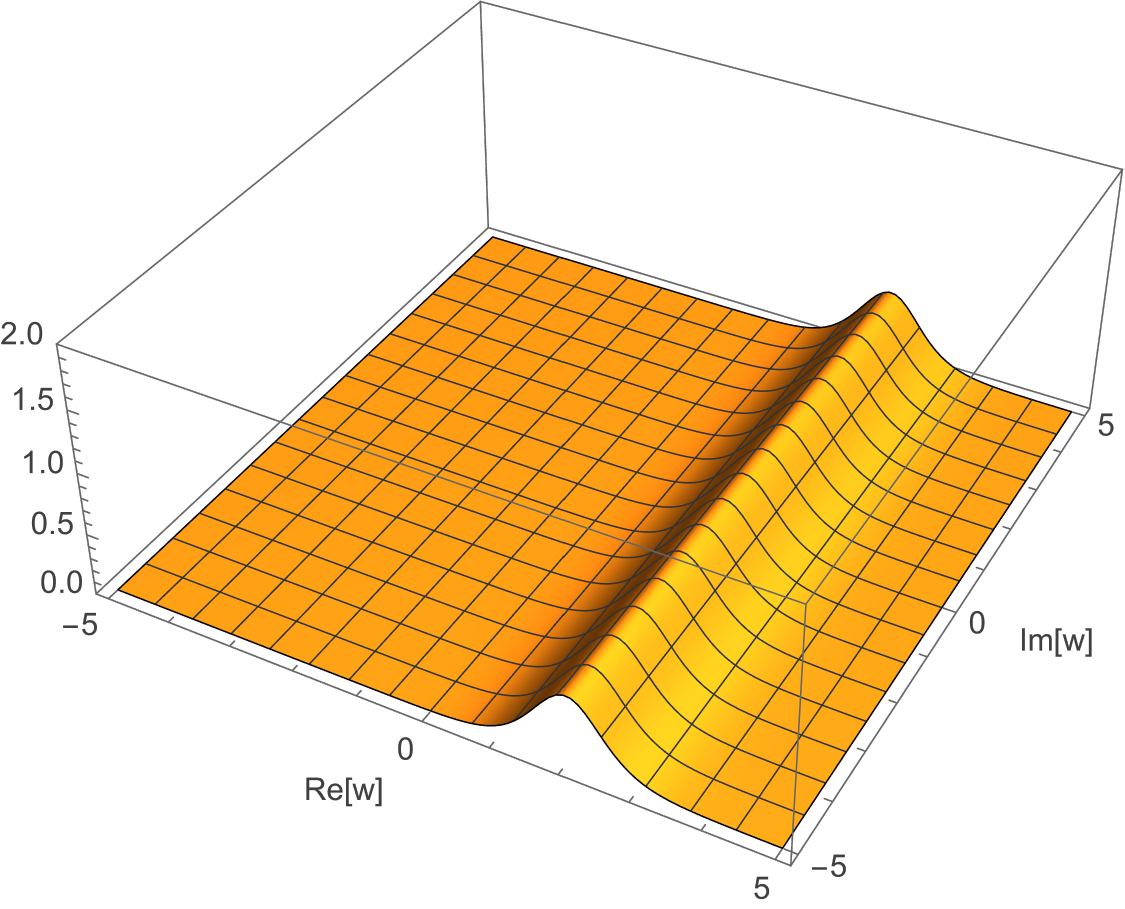}
\caption{$f(w) = \exp(\frac{3}{2} (w - 2))$.}
\end{subfigure}%
\vspace{1.5em}\\%
\begin{subfigure}{0.45\textwidth}
\centering
\includegraphics[width=\textwidth]{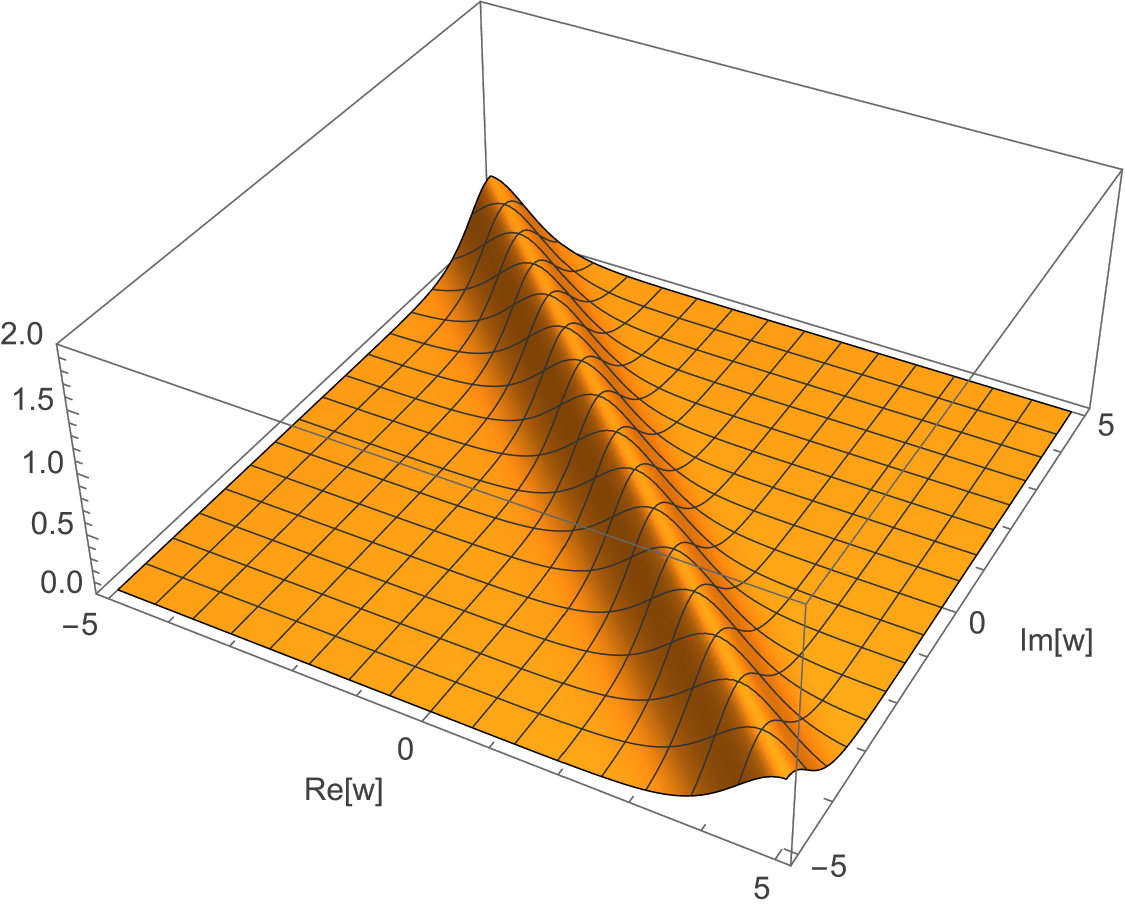}
\caption{$f(w) = \exp(\frac{3}{2} e^{-\frac{\iu \pi}{4}} w)$.}
\end{subfigure}%
\hfill
\begin{subfigure}{0.45\textwidth}
\centering
\includegraphics[width=\textwidth]{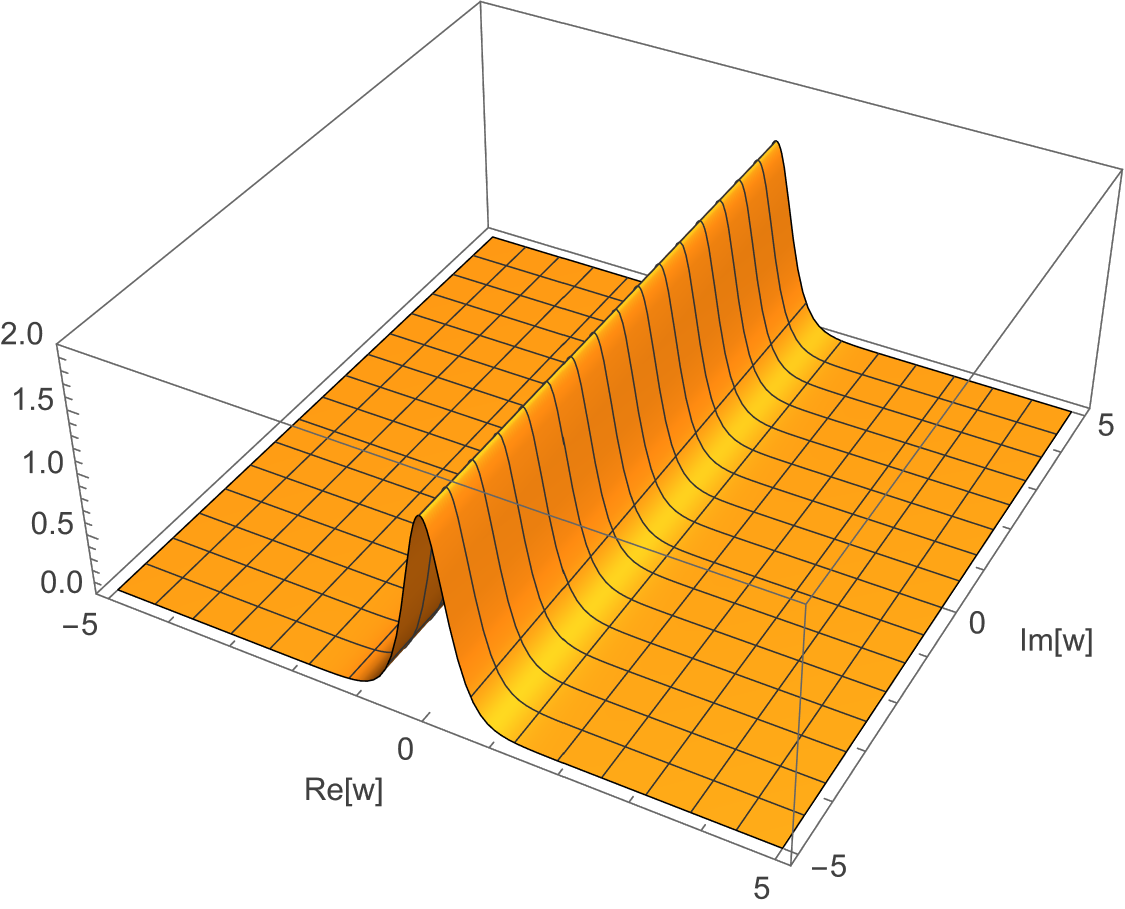}
\caption{$f(w) = \exp(\frac{5}{2} w)$.}
\end{subfigure}%
\vspace{1em}
\caption{
    Exponential maps of the form $f(w) = \exp(s(w))$ correspond to line solitons,
    where $s(w) = c \, (e^{\iu \psi} w - \protect\rchi_0)$.
    The parameters control translations ($\protect\rchi_0$), rotations ($\psi$),
    and rescalings ($c$) of the $w$-plane, as well as the height of the line ($c$).
    The energy density is proportional to $\frac{\vert f^\prime \vert^2}{(1 + \vert f \vert^2)^2}$,
    which has been plotted over the $w$-plane for various examples.
}
\label{fig:lines}
\end{figure}

In addition to lump solitons, the 3d ICM admits extended line solitons, as shown in~\cite{Lee89}.
Unlike the lump solitons, the energy density of these extended solutions
does not tend to zero as $\vert w \vert \to \infty$, meaning that their total energy diverges.
The associated maps $f : \fC \to \CP^1$ do not have finite degree, and their image wraps
the target space $\CP^1$ infinitely many times.
A simple example of an extended solution is given by the map $f(w) = \exp(w)$,
whose energy density is proportional to
\begin{equation}
    \frac{\pd_w f \pd_{\bar{w}} \bar{f}}{(1 + \vert f \vert^2)^2} =
    \frac{\exp(w + \bar{w})}{(1 + \exp(w + \bar{w}))^2} =
    \frac{1}{4} \sech(\Re(w))^2 \,.
\end{equation}
This energy density is constant along the imaginary axis and peaked at $\Re(w) = 0$,
allowing for its interpretation as a line soliton lying along $\Re(w) = 0$.
The location of the line soliton in the $w$-plane can be adjusted by considering maps
of the form $f(w) = \exp(s(w))$ where $s(w) = c \, (e^{\iu \psi} w - \rchi_0)$,
which allows for translations ($\rchi_0$), rotations ($\psi$), and rescalings ($c$) of the $w$-plane.
The energy density of these line solitons is proportional to
\begin{equation}
    \frac{\pd_w f \pd_{\bar{w}} \bar{f}}{(1 + \vert f \vert^2)^2} =
    \frac{\vert c \vert^2}{4} \sech(\Re(s))^2 \,.
\end{equation}
The line soliton now lies along $\Re(s) = 0$, and its height is proportional to $\vert c \vert^2$.
Some examples are presented in~\figref{fig:lines}.

\begin{figure}
\begin{subfigure}{0.45\textwidth}
\centering
\includegraphics[width=\textwidth]{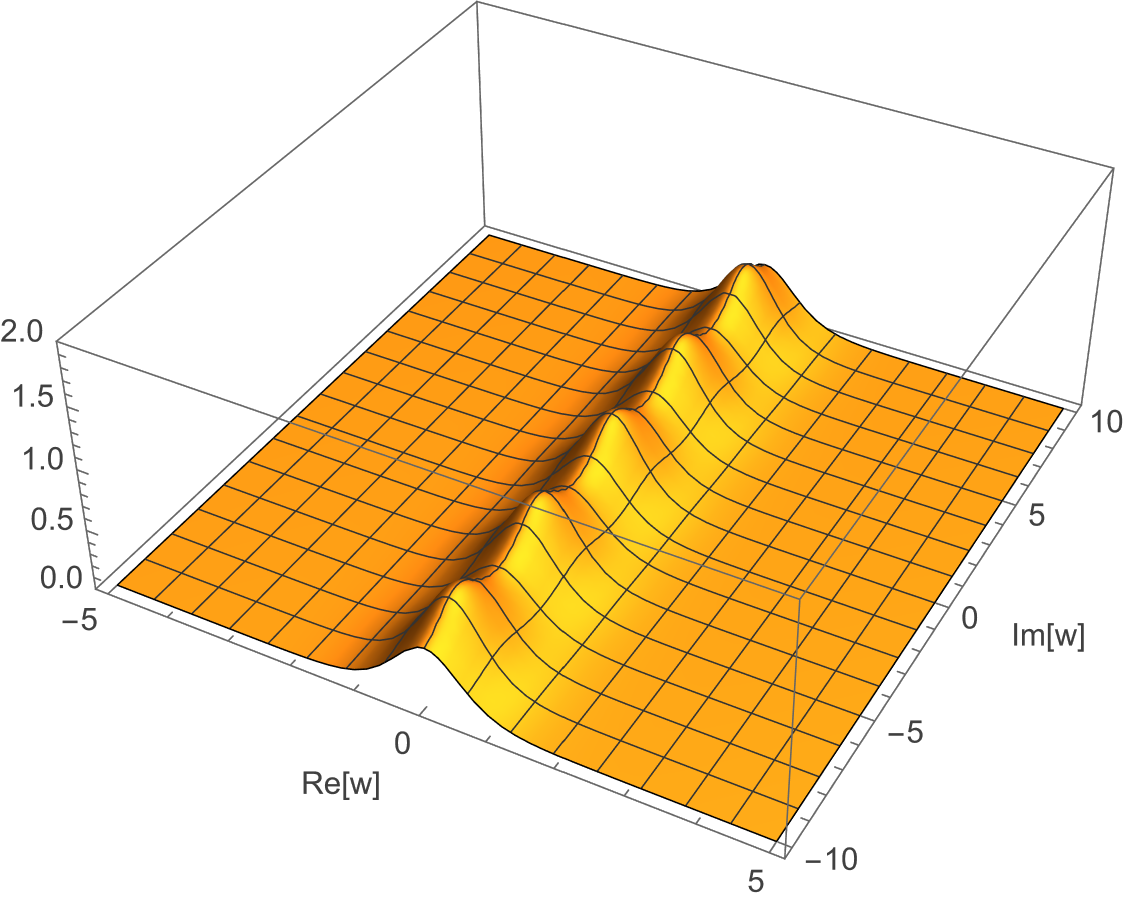}
\caption{$f(w) = \exp(\frac{3}{2} w) - \frac{1}{10}$.}
\end{subfigure}%
\hfill
\begin{subfigure}{0.45\textwidth}
\centering
\includegraphics[width=\textwidth]{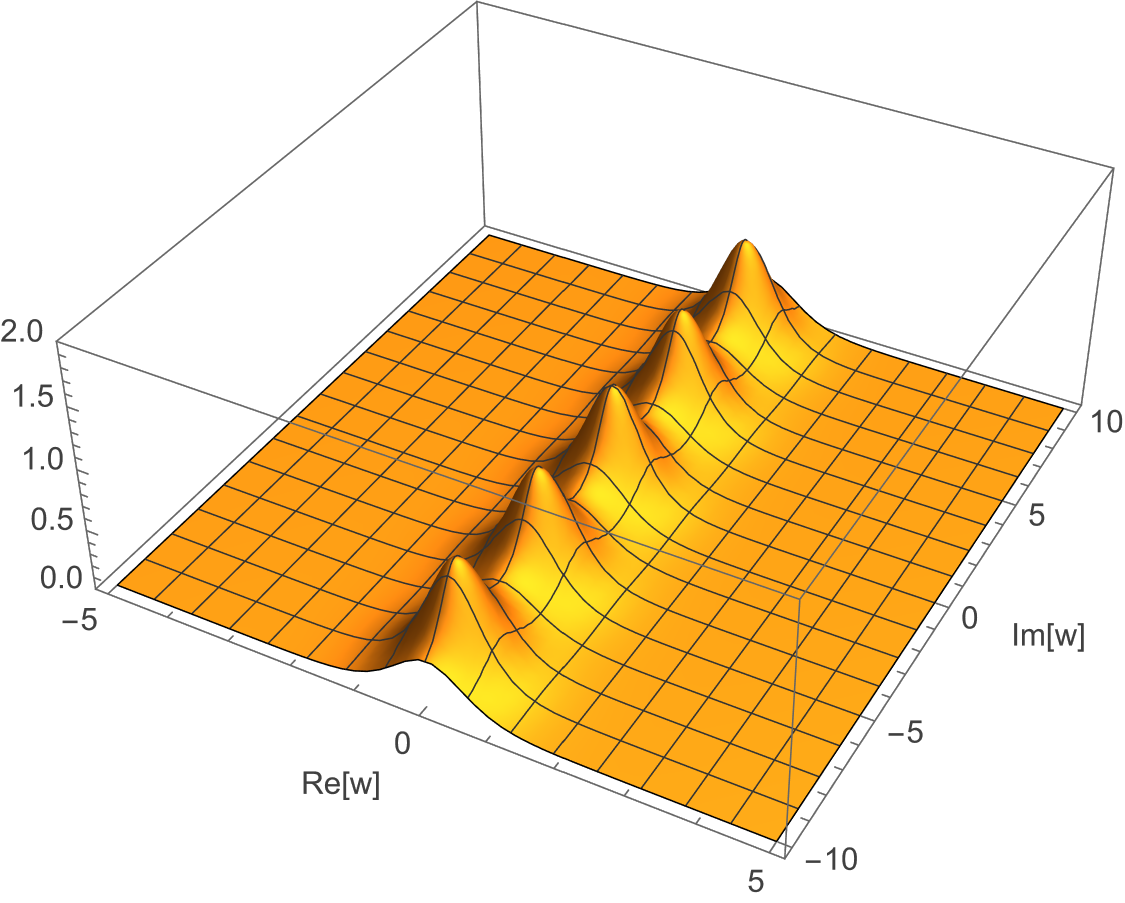}
\caption{$f(w) = \exp(\frac{3}{2} w) - \frac{1}{4}$.}
\end{subfigure}%
\vspace{1.5em}\\%
\begin{subfigure}{0.45\textwidth}
\centering
\includegraphics[width=\textwidth]{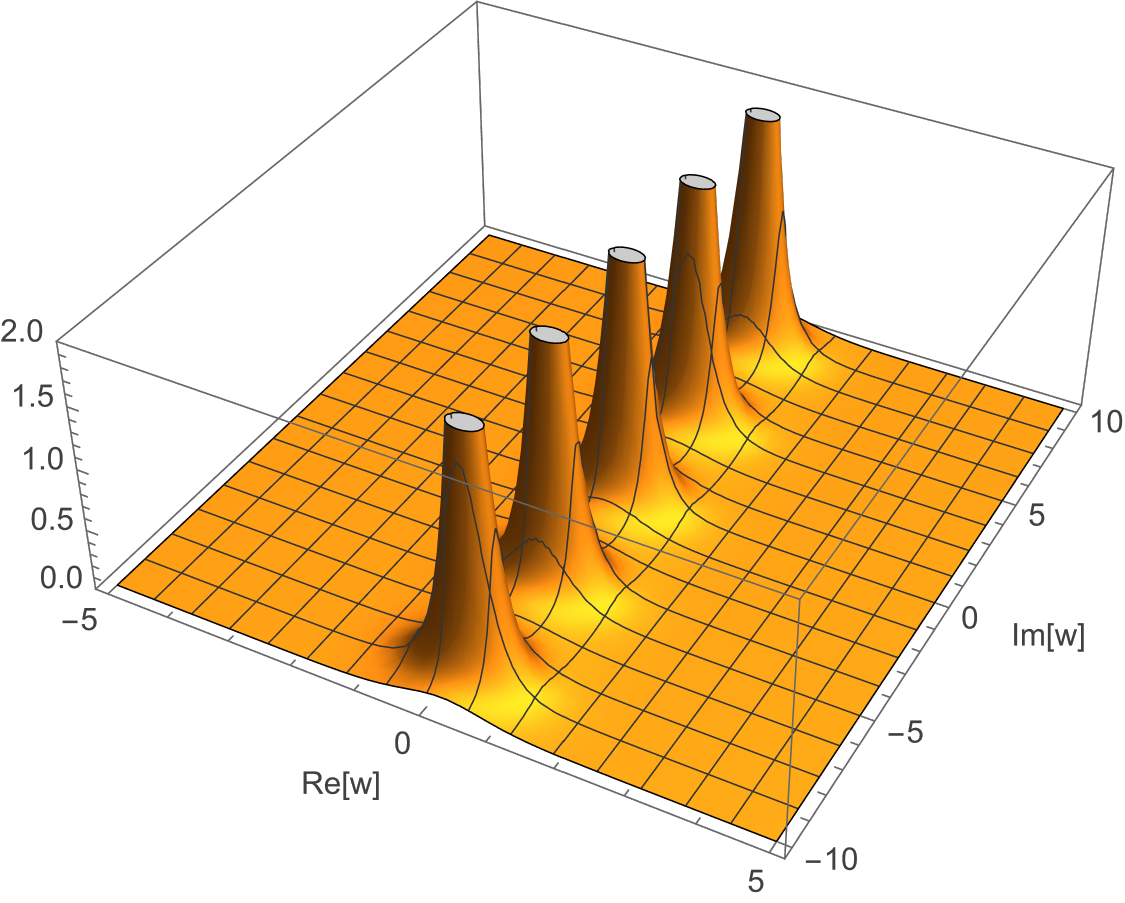}
\caption{$f(w) = \exp(\frac{3}{2} w) - 1$.}
\end{subfigure}%
\hfill
\begin{subfigure}{0.45\textwidth}
\centering
\includegraphics[width=\textwidth]{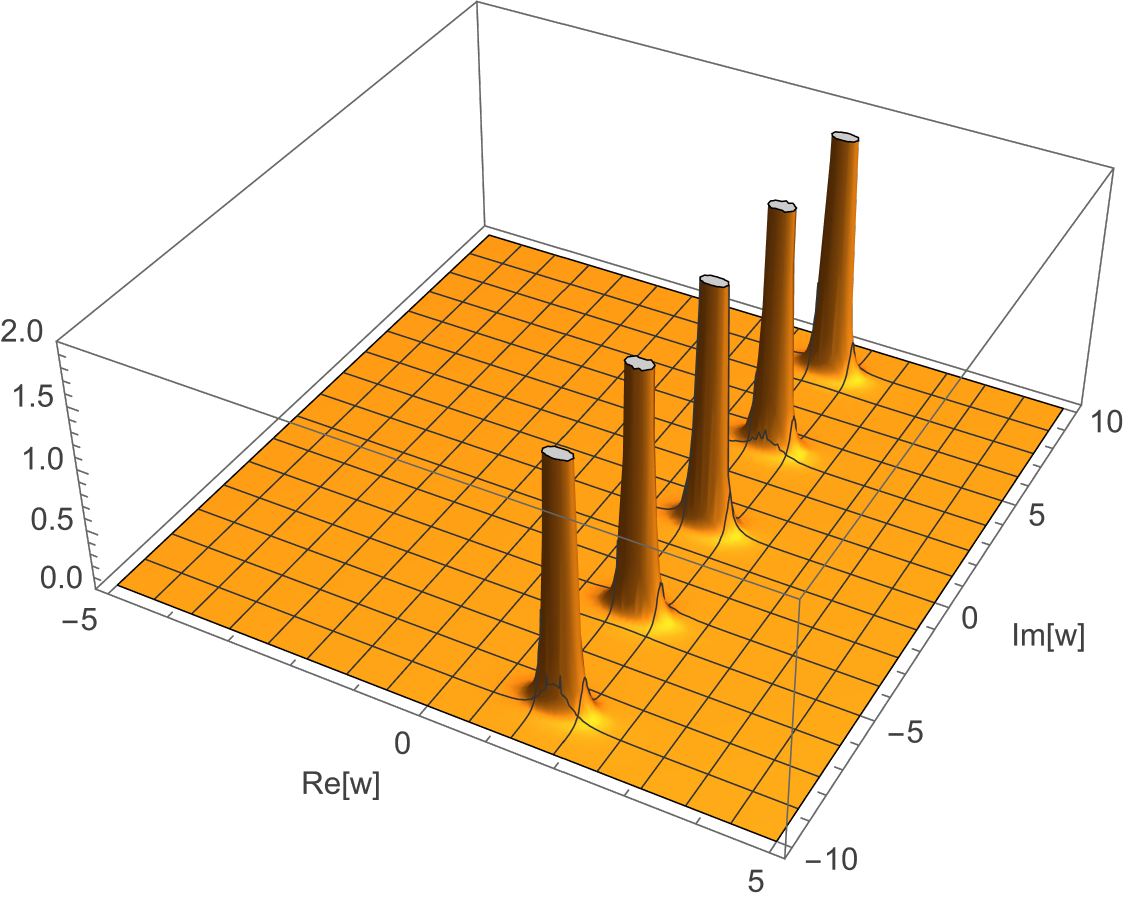}
\caption{$f(w) = \exp(\frac{3}{2} w) - 10$.}
\end{subfigure}%
\vspace{1em}
\caption{
    Exponential maps of the form $f(w) = \exp(s(w)) + Z_0$ describe domain walls which tend towards
    $Z \to \infty$ as $\Re(s) \to +\infty$ and $Z \to Z_0$ as $\Re(s) \to -\infty$.
    If $\vert Z_0 \vert^2$ is relatively small, the energy density still looks like a line soliton,
    but the line now has an oscillating profile along the tangential direction.
    As $\vert Z_0 \vert^2$ becomes larger, this oscillating profile dominates, leaving
    a set of disconnected peaks separated by $s \mapsto s + 2 \pi \iu$.
    The energy density is proportional to $\frac{\vert f^\prime \vert^2}{(1 + \vert f \vert^2)^2}$,
    which has been plotted over the $w$-plane for various examples.
}
\label{fig:lines_vacua}
\end{figure}

\medskip

Let us consider the movement of these line solitons in spacetime.
At any fixed time, the equation $\Re(s) = 0$ defines a line in the $xy$-plane
whose normal vector is given by
\begin{equation}
    n_x = \cos(\psi) - r^2 \cos(\psi + 2 \theta) \,, \qquad
    n_y = 2 \, r \cos(\psi + \theta) \,, \qquad
    \hat{n} = \frac{n}{\Vert n \Vert} \,.
\end{equation}
The velocity ($v$) of the lump soliton computed earlier can equivalently be thought of
as the velocity of the $w$-plane relative to the $xy$-plane,
and this vector generically has a tangential and normal component relative to the line soliton.
There are two extreme circumstances: the $w$-plane velocity is completely tangent to the line,
or the $w$-plane velocity is completely normal to the line.
These occur when the following conditions are satisfied~\cite{Lee89}.
\begin{equation}\begin{aligned}
    \hat{n} \cdot v & = 0 \quad & & \iff & \quad
    (-1 + r^2) \tan(\psi + \theta) & = (1 + r^2) \cot(\theta) \,, \\
    \hat{n} \cdot v & = \Vert v \Vert \quad & & \iff & \quad
    (1 + r^2) \tan(\psi + \theta) & = (1 - r^2) \tan(\theta) \,.
\end{aligned}\end{equation}
Therefore, at fixed $\alpha = r e^{\iu \theta}$, the normal velocity of the line
varies between zero and $\Vert v \Vert$, depending on its orientation in the $w$-plane.
In particular, the line solitons can travel at any velocity below the speed of light.

\medskip

The line solitons also admit an interpretation as domain walls,
interpolating between two vacua of the theory.
For fixed time, the line $\Re(s) = 0$ divides the $xy$-plane into two regions,
given by $\Re(s) < 0$ and $\Re(s) > 0$ respectively.
Let us compare the asymptotic field configuration as $\Re(s)$ tends to either
negative or positive infinity.
As $\Re(s) \to -\infty$, the map $f = \exp(s)$ tends to zero,
meaning that the field configuration tends to $Z \to 0$.
By comparison, as $\Re(s) \to +\infty$, the field configuration tends to $Z \to \infty$.
Since the 3d ICM has no potential term, any fixed group element is a valid vacuum,
and the line soliton solution can be understood as interpolating between these two different vacua.

\medskip

This interpretation prompts us to consider the generalisation $f(w) = \exp(s(w)) + Z_0$,
which now tends to $Z \to Z_0$ as $\Re(s) \to -\infty$.
The energy density of these solutions is proportional to
\begin{equation}
    \frac{\vert f^\prime \vert^2}{(1 + \vert f \vert^2)^2} =
    \frac{\vert s^\prime \vert^2 \, \vert \exp(s) \vert^2}{(1 + \vert \exp(s) + Z_0 \vert^2)^2} \,.
\end{equation}
If $\vert Z_0 \vert^2$ is relatively small, the energy density still looks like a line soliton,
but the line now has an oscillating profile along the tangential direction.
As $\vert Z_0 \vert^2$ becomes larger, this oscillating profile dominates, leaving
a set of disconnected peaks separated by $s \mapsto s + 2 \pi \iu$.
Some examples are presented in~\figref{fig:lines_vacua}.

\section{Soliton Scattering}\label{sec:scattering}
Let us set up a scattering experiment between these solitons
in the 3d integrable chiral model.
Classically, scattering can be understood by comparing the field configurations
in the far past ($t \to -\infty$) and in the far future ($t \to +\infty$).
In these limits, the solitons will generically be well-separated,
and the solution can be approximated by a superposition of $1$-soliton solutions.
The parameters associated with a $1$-soliton solution are a point $\alpha \in \CP^1$
and a holomorphic map $f \in \cat{Hol}(\fC, \CP^1$).
We will study soliton scattering by comparing these parameters in the far past and the far future,
with a change in these parameters corresponding to a nontrivial scattering event.

\medskip

Both of the original papers on the 3d integrable chiral model~\cite{MZ81,War88}
highlight that generic lump solitons do not scatter off each other.
\unskip\footnote{
While generic lump solitons do not scatter off each other,
nontrivial lump scattering does occur after
taking a limit in which the poles of the dressing transformations coincide~\cite{War95,Ioa96,DT07}.
}
Scattering of extended objects in the 3d ICM was first studied in~\cite{Lee89},
where it was shown that line solitons interact nontrivially
with lump solitons and with other line solitons.
Also, line soliton scattering in the 3d ICM on noncommutative spacetime
was studied in~\cite{Bie02}.
Related work studied line soliton scattering in the 2+2d Wess-Zumino-Witten model~\cite{LHHZ25}.

\subsection{\texorpdfstring{$2 \to 2$}{2 to 2} Interactions}
A $2 \to 2$ scattering process corresponds to a $2$-soliton dressing transformation,
which can be constructed using the parameters
$\{ \alpha_1 , \hat{a}_1 = (1, \hat{f}_1) \}$ and $\{ \alpha_2 , \hat{a}_2 = (1, \hat{f}_2) \}$.
It is useful to write this $2$-soliton solution as the product of two $1$-soliton transformations.
Firstly, applying the $1$-soliton transformation with parameters
$\{ \alpha_i , \hat{a}_i = (1, \hat{f}_i) \}$ gives
\begin{equation}
    U_i = \id - \frac{\alpha_i - \bar\alpha_i}{z - \bar\alpha_i} \, P(\hat{a}_i) \,, \qquad
    P(\hat{a}_i) = \frac{\hat{a}_i \, \hat{a}_i^\dagger}{\Vert \hat{a}_i \Vert^2} \,.
\end{equation}
Then, a subsequent $1$-soliton transformation with parameters
$\{ \alpha_j , \hat{a}_j = (1, \hat{f}_j) \}$ gives
\begin{equation}
    U_{ij} = H_{ij}^{-1} U_i \,, \qquad
    H_{ij}^{-1} = \id - \frac{\alpha_j - \bar\alpha_j}{z - \bar\alpha_j} \, P(a_{ij}) \,, \qquad
    a_{ij} = (U_i \vert_{\alpha_j}) \, \hat{a}_j \,.
\end{equation}
It is important to emphasise that the order of these $1$-soliton transformations does not matter,
meaning that the $2$-soliton solution is symmetric in its indices $U_{ij} = U_{ji}$.
One way to demonstrate this fact is to show that both matrices can be written as
\begin{equation}\begin{aligned}
    U_{ij} & = \id + \frac{Y_i \, \hat{a}_i^\dagger}{z - \bar\alpha_i} + \frac{Y_j \, \hat{a}_j^\dagger}{z - \bar\alpha_j} \,, \quad & \quad
    U_{ij}^{-1} & = \id + \frac{\hat{a}_i \, X_i}{z - \alpha_i} + \frac{\hat{a}_j \, X_j}{z - \alpha_j} \,, \\
    U_{ji} & = \id + \frac{\tilde{Y}_i \, \hat{a}_i^\dagger}{z - \bar\alpha_i} + \frac{\tilde{Y}_j \, \hat{a}_j^\dagger}{z - \bar\alpha_j} \,, \quad & \quad
    U_{ji}^{-1} & = \id + \frac{\hat{a}_i \, \tilde{X}_i}{z - \alpha_i} + \frac{\hat{a}_j \, \tilde{X}_j}{z - \alpha_j} \,.
\end{aligned}\end{equation}
Once this has been established, the unknowns
$\{ X_i, X_j, Y_i, Y_j, \tilde{X}_i, \tilde{X}_j, \tilde{Y}_i, \tilde{Y}_j \}$ are
completely determined by the conditions $U_{ij}^{-1} U_{ij} = \id$ and $U_{ji}^{-1} U_{ji} = \id$.
Therefore, showing that both matrices can be written in this form implies that $U_{ij} = U_{ji}$.

\medskip

By construction, $U_{ij}$ and $U_{ji}$ are meromorphic matrices which
tend to the identity matrix as $z \to \infty$.
Therefore, showing that they can be written in the form above amounts to showing
that their residues take the required form.
For example, the residues at $z = \bar\alpha_i$ can be written as
\begin{equation}\begin{aligned}
    \res_{\bar\alpha_i} (U_{ij}) & = - (\alpha_i - \bar\alpha_i) (H_{ij}^{-1} \vert_{\bar\alpha_i}) \, P(\hat{a}_i)
    = Y_i \, \hat{a}_i^\dagger \,, \\
    \res_{\bar\alpha_i} (U_{ji}) & = - (\alpha_i - \bar\alpha_i) \, P(a_{ji}) \, (U_j \vert_{\bar\alpha_i})
    = \tilde{Y}_i \, \hat{a}_i^\dagger \,.
\end{aligned}\end{equation}
A similar argument holds for the other residues,
allowing us to conclude that the $2$-soliton solution
is symmetric in its indices $U_{ij} = H_{ij}^{-1} U_i = H_{ji}^{-1} U_j = U_{ji}$.

\medskip

In both asymptotic regions $(t \to \pm \infty)$, the $2$-soliton solution
is well approximated by a superposition of two $1$-soliton solutions.
The parameters of these $1$-soliton solutions can be found by restricting the $2$-soliton solution
to a small region around each $1$-soliton.
For example, in a small region around the $1$-soliton associated with $\alpha_j$,
the map $\hat{f}_i$ can be approximated by some fixed value $\hat{f}_i = Z_i$.
In this case, $U_i$ depends on $z$ but does not depend on spacetime,
and all of the spacetime-dependence in $U_{ij}$ is contained in $\hat{f}_j$.

\medskip

The $z$-dependent matrix $U_i$ plays two roles in the $2$-soliton solution $U_{ij}$.
It appears explicitly as an overall right-multiplication, which amounts to
a global $\grp{U}(2)$ symmetry transformation after evaluating at $z = 0$,
and it contributes to the profile of the $1$-soliton associated with $\alpha_j$
through the expression $a_{ij} = (U_i \vert_{\alpha_j}) \, \hat{a}_j$.
While the reality conditions ensure that $U_i$ is valued in $\grp{U}(2)$ when evaluated
at real values of $z$, it is not required to lie in any particular subspace of
$\grp{GL}(2, \fC)$ when evaluated at $z = \alpha_j$.
Therefore, the matrix $U_i \vert_{\alpha_j}$ generically implements a $\grp{GL}(2, \fC)$
transformation on the $1$-soliton parameters $\hat{a}_j = (1, \hat{f}_j)$,
which reduces to an $\grp{SL}(2, \fC)$ transformation due to the rescaling symmetry.
This $\grp{SL}(2, \fC)$ action can be identified with
the M{\" o}bius transformations of the target space $\CP^1$,
and only the $\grp{SU}(2)$ subgroup of these transformations preserves the energy density,
so these transformations will generically modify the profile of the $1$-soliton.

\medskip

For example, consider a scattering event between two $1$-lump solitons defined by
\begin{equation}
    \hat{f}_1 (w) = c_1 (w - \rchi_1) \,, \qquad
    \hat{f}_2 (w) = c_2 (w - \rchi_2) \,.
\end{equation}
In both asymptotic regions $(t \to \pm \infty)$, these functions tend to $\hat{f}_i \to \infty$
at generic points in the $xy$-plane.
The exceptions are the locations of the two $1$-lump solitons, determined by the equations
$w_{\alpha_1} = \rchi_1$ and $w_{\alpha_2} = \rchi_2$ respectively.
Let us focus on a small region around the $1$-lump soliton associated with $\alpha_j$,
where the $z$-dependent matrix $U_i$ is well approximated by
\begin{equation}
    U_i = 
    \begin{pmatrix}
        1   & 0 \\
        0   & \frac{z - \alpha_i}{z - \bar\alpha_i}
    \end{pmatrix} \,.
\end{equation}
Evaluating this at $z = 0$ gives the global $\grp{U}(2)$ action explicitly appearing in $U_{ij}$,
and the $1$-soliton profile in this region is determined by
\begin{equation}
    a_{ij} = (U_i \vert_{\alpha_j}) \, \hat{a}_j
    = (1, f_{ij}) \,, \qquad
    f_{ij} = \frac{\alpha_j - \alpha_i}{\alpha_j - \bar\alpha_i} \, \hat{f}_j \,.
\end{equation}
This means that the effective heights of the $1$-lump solitons depend not only on the parameters
$c_i$ in the maps $\hat{f}_i$, but also on the relative locations of the poles $\alpha_i$.
On the other hand, the trajectories of the $1$-lump solitons are completely fixed by
the poles $\alpha_i$ and the centres of mass $\rchi_i$.
These are not statements about the effect of any scattering process,
but rather statements about the relationship between the effective physical parameters and
the input parameters in our solution generating technique.
This relationship is important to bear in mind when considering the forthcoming analysis.
We will continue to write expressions in terms of the input parameters
as these are more readily accessible, and a dictionary would be required
in order to write these expressions in terms of the physical parameters of the 3d solution.
This dictionary can be constructed by studying the asymptotic solution directly,
as we have just shown in this simple example.

\medskip

Scattering can be studied by comparing the physical parameters
in the far past ($t \to -\infty$) and in the far future ($t \to +\infty$).
In particular, for lump solitons the analysis is unchanged between the two regions
because the limit of the functions $\hat{f}_i \to \infty$ is the same in both cases.
In other words, the effective parameters of the $1$-soliton solutions do not differ
between the two asymptotic regions, meaning that the scattering is trivial.
Furthermore, this result holds whenever
$\hat{f}_1$ and $\hat{f}_2$ describe solitons with finite energy.
All of these maps extend from $\fC$ to $\CP^1$~\cite{Uhl89,War90},
meaning that they are single-valued as $\vert w \vert \to \infty$
and therefore have trivial scattering.
By comparison, line solitons correspond to exponential maps
which have different asymptotic values as $\vert w \vert \to \infty$,
so they still stand a chance of exhibiting nontrivial scattering.

\medskip

Consider a scattering event between a line soliton and a lump soliton defined by
\begin{equation}
    \hat{f}_1 (w) = \exp(s_1(w)) + Z_1 \,, \qquad
    \hat{f}_2 (w) = c_2 (w - \rchi_2) \,,
\end{equation}
where $s_1(w) = c_1 (e^{\iu \psi_1} w - \rchi_1)$.
Far away from the lump soliton at $w_{\alpha_2} = \rchi_2$, the line soliton can be approximated
by a $1$-soliton solution with parameters
$\{ \alpha_1, f_{21} = \frac{\alpha_1 - \alpha_2}{\alpha_1 - \bar\alpha_2} \hat{f}_1 \}$.
This is the same in both asymptotic regions, so the line soliton is not modified by the scattering.
At any fixed time, the line soliton splits the $xy$-plane into two domains whose asymptotic regions
are $\Re(s_1) \to -\infty$ and $\Re(s_1) \to +\infty$ respectively.
Generically, the trajectory of the lump soliton will pass through the line soliton
at some finite time, meaning that it will pass from one domain in the $xy$-plane to the other.
The two asymptotic profiles for the lump soliton can therefore be found by taking the limits
$\Re(s_1) \to -\infty$ and $\Re(s_1) \to +\infty$ in the $2$-soliton solution,
which are equivalent to taking $\hat{f}_1 \to Z_1$ and $\hat{f}_1 \to \infty$ respectively.

\begin{figure}
\centering
\includegraphics[width=\textwidth]{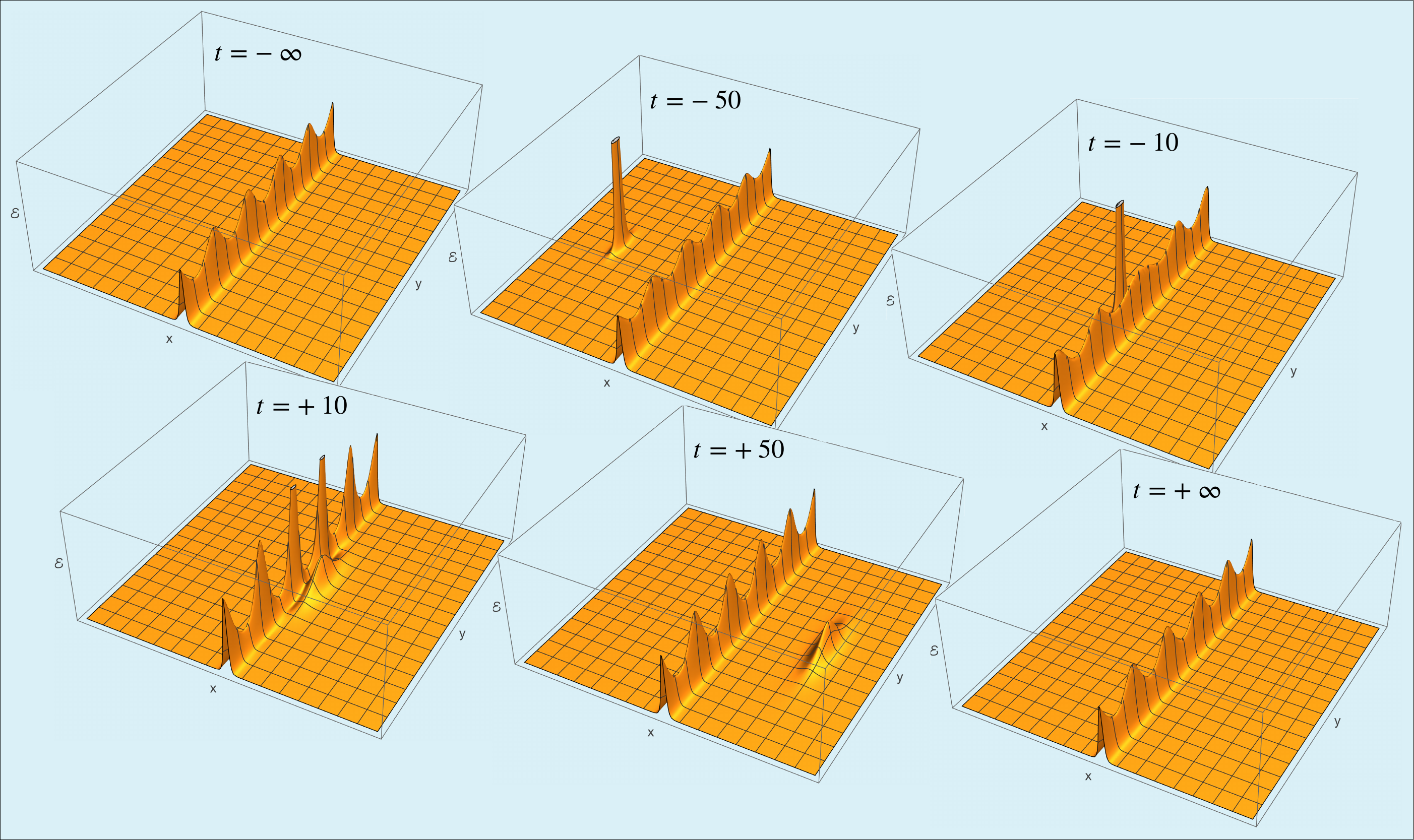}
\caption{
    A scattering event between a line soliton and a lump soliton defined by
    $\{ \alpha_1 = \iu, \hat{f}_1 (w) = \exp(w) + \frac{1}{2} \}$ and
    $\{ \alpha_2 = 2 \iu, \hat{f}_2 (w) = \frac{w}{2} \}$.
    The energy density is plotted over space at various fixed times.
}
\label{fig:linelumpscattering}
\end{figure}

\medskip

In the region $\Re(s_1) \to +\infty$,
the lump soliton is well approximated by the same $1$-soliton solution as before,
since the line soliton is well approximated by $\hat{a}_1 = (0,1)$.
On the other hand, in the region $\Re(s_1) \to -\infty$,
the matrix $U_1$ is well approximated by
\begin{equation}
    U_1 = \id - \frac{\alpha_1 - \bar\alpha_1}{z - \bar\alpha_1} \, P(\hat{a}_1) \,, \qquad
    \hat{a}_1 = (1, Z_1) \,.
\end{equation}
Let us introduce the notation $U_i(z; Z)$ for the $z$-dependent matrix $U_i$ evaluated on
the fixed configuration $\hat{f}_i = Z$.
In this notation, the lump soliton is well approximated by the profile
$a_{12} = U_1(\alpha_2; \infty) \, \hat{a}_2$ in the region $\Re(s_1) \to +\infty$,
and it is well approximated by the profile
$a_{12} = U_1(\alpha_2; Z_1) \, \hat{a}_2$ in the region $\Re(s_1) \to -\infty$.
Therefore, the net effect of the scattering on the lump soliton profile is
the $\grp{SL}(2, \fC)$ transformation
\begin{equation}
    a_{12} \mapsto U_1(\alpha_2; Z_1) U_1(\alpha_2; \infty)^{-1} a_{12} \,.
\end{equation}
The global $\grp{U}(2)$ symmetry transformation $U_1(0; \infty)^{-1} U_1(0; Z_1)$
should also be included to capture the full scattering process.

\medskip

In order to provide a physical interpretation for this scattering process,
let us study the energy density of the lump soliton before and after the scattering event.
Since we are restricting our attention to the energy density, we can ignore
the global $\grp{U}(2)$ symmetry transformation as this leaves the energy density invariant.
Furthermore, the $\grp{SU}(2)$ subgroup of $\grp{SL}(2, \fC)$ also preserves the energy density,
and this component can be extracted by performing an Iwasawa decomposition of $\grp{SL}(2, \fC)$.
Up to an $\grp{SU}(2)$ transformation $M \in \grp{SU}(2)$ and
an overall rescaling $\rho \in \fC^\ast$,
the $\grp{SL}(2, \fC)$ transformation can be written as
\begin{equation}\begin{gathered}\label{eq:scatteringtransformation}
    U_1(\alpha_2; Z_1) U_1(\alpha_2; \infty)^{-1} = \rho \, M
    \begin{pmatrix}
        1           & 0 \\
        \mathcal{B} & \mathcal{A}
    \end{pmatrix} \,, \\
    \mathcal{A} = \frac{
        \vert \alpha_1 - \bar\alpha_2 \vert^2 + \vert Z_1 \vert^2 \vert \alpha_1 - \alpha_2 \vert^2
    }{
        (1 + \vert Z_1 \vert^2) \vert \alpha_1 - \alpha_2 \vert^2
    } \,, \qquad
    \mathcal{B} = \frac{
        Z_1 (\alpha_1 - \bar\alpha_1) (\alpha_2 - \bar\alpha_2)
    }{
        (1 + \vert Z_1 \vert^2) (\bar\alpha_1 - \alpha_2) (\bar\alpha_1 - \bar\alpha_2)
    } \,.
\end{gathered}\end{equation}
Therefore, the change in the energy density of the lump soliton is captured by the map
\begin{equation}
    f_{12} \mapsto \mathcal{A} \, f_{12} + \mathcal{B} \,.
\end{equation}
We can interpret this transformation in terms of the physical parameters of the lump soliton.
The height of the lump soliton $c$ will transform as $c \mapsto \mathcal{A} \, c$,
and the centre of mass $\rchi_0$ transforms as
$\rchi_0 \mapsto \rchi_0 - \frac{\mathcal{B}}{\mathcal{A} c}$.
The shift in the centre of mass is generically complex,
so some component will point along the velocity vector resulting in a time delay,
and some component will be normal to the velocity resulting in a shift of the overall trajectory.
We conclude that the lump solitons experience nontrivial scattering
when interacting with line solitons.
An example scattering event between a line soliton and a lump soliton
is presented in~\figref{fig:linelumpscattering}.

\medskip

As the final $2 \to 2$ process, consider a $2$-soliton solution for two lines defined by
\begin{equation}
    \hat{f}_1 (w) = \exp(s_1(w)) + Z_1 \,, \qquad
    \hat{f}_2 (w) = \exp(s_2(w)) + Z_2 \,,
\end{equation}
where $s_1(w) = c_1 (e^{\iu \psi_1} w - \rchi_1)$ and $s_2(w) = c_2 (e^{\iu \psi_2} w -\rchi_2)$.
Before studying the $1$-soliton profiles in various regions,
let us consider the spacetime geometry.
The worldvolume of each line soliton defines a 2d plane in 3d spacetime,
and these will generically intersect along a 1d line.
This 1d line will generically intersect each spatial slice at a single point,
and this point will move through space as the solution evolves with time.
In this circumstance, there is no particular time at which a scattering event takes place;
rather, there is an $X$-shaped soliton profile which moves through space.
A third line soliton is required for there to be a scattering event,
because the intersection of three 2d planes is generically a point in 3d spacetime.
However, in the case that the two line solitons are parallel in space,
their 2d worldvolumes will intersect at some finite fixed time.
In this circumstance, it is reasonable to talk about a scattering event between two lines.

\begin{figure}
\centering
\includegraphics[width=\textwidth]{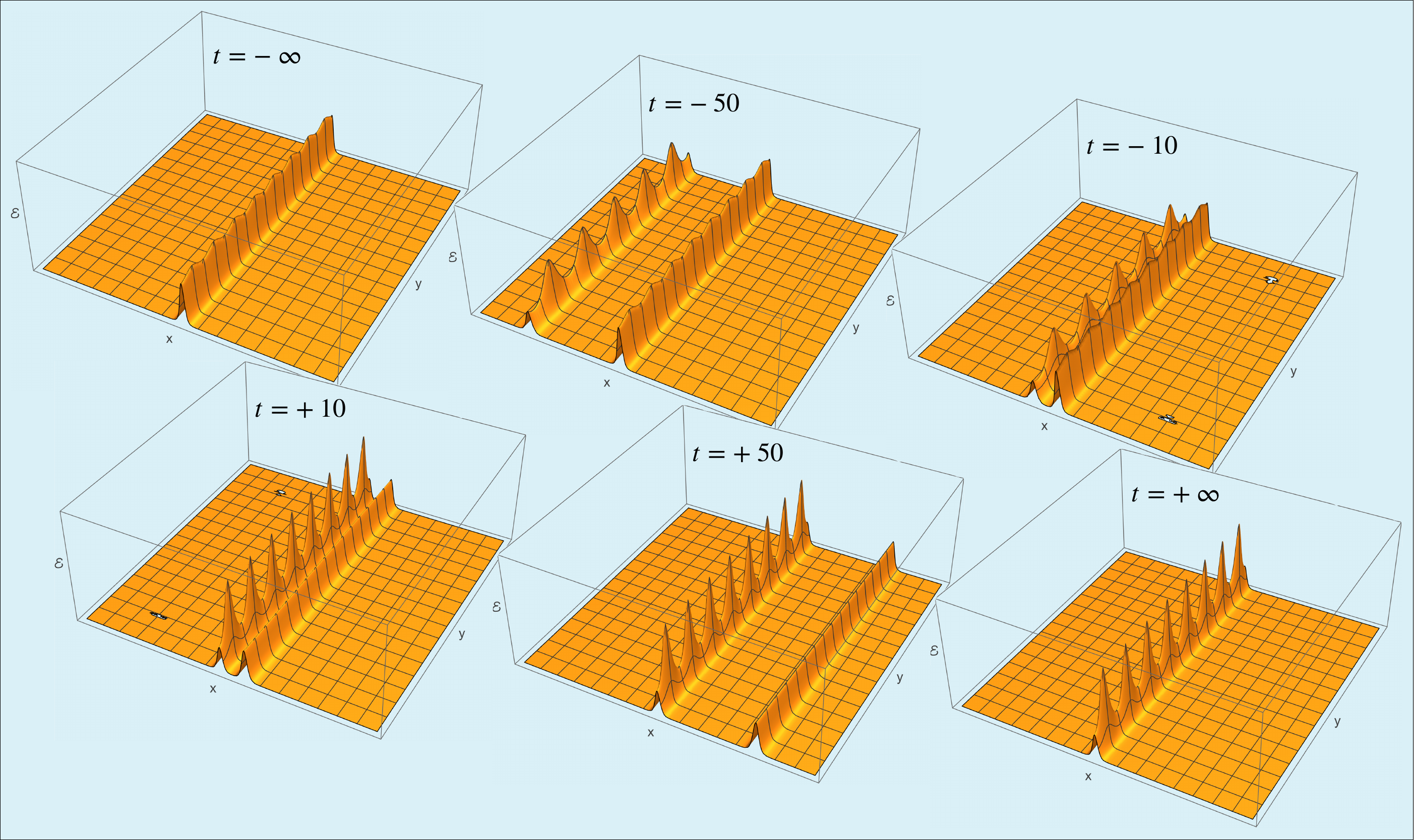}
\caption{
    A scattering event between two parallel line solitons that are defined by
    $\{ \alpha_1 = \iu, \hat{f}_1 (w) = \exp(w) + \frac{1}{10} \}$ and
    $\{ \alpha_2 = 2 \iu, \hat{f}_2 (w) = \exp (\frac{w}{3}) \}$.
    The energy density is plotted over space at fixed times.
}
\label{fig:linelinescattering}
\end{figure}

\medskip

While the physical interpretation will differ between these two cases,
in either case it is useful to study the $1$-soliton profiles in various regions.
For parallel lines, this will describe a nontrivial scattering event
which relates the profile in the far past to the profile in the far future.
For the $X$-shaped soliton, each line is split into two components by the point of intersection,
and this analysis will relate the profiles along these two components.
The profile of the line soliton associated with $\alpha_2$ in the region $\Re(s_1) \to +\infty$
is related to its profile in the region $\Re(s_1) \to -\infty$ by the same map as before,
$f_{12} \mapsto \mathcal{A} \, f_{12} + \mathcal{B}$.
Conversely, the inverse transformation is given by
\begin{equation}
    f_{12} \mapsto \mathcal{A}^{-1}  (f_{12} - \mathcal{B}) \,.
\end{equation}
Similarly, the transformation of the line soliton associated with $\alpha_1$ can be found by
exchanging the indices in the previous expressions.
Let us give a physical interpretation for the line soliton associated with $\alpha_2$
under the transformation $f_{12} \mapsto \mathcal{A} \, f_{12} + \mathcal{B}$.
The function $s(w)$ transforms as $s(w) \mapsto s(w) + \log(\mathcal{A})$,
and this can be absorbed into a real shift $\rchi_0 \mapsto \rchi_0 - \log(\mathcal{A}) / c$
which results in a time delay.
In addition, the asymptotic value $Z_0$ transforms as $Z_0 \mapsto Z_0 + \mathcal{B}$.
Analogous expressions hold for the other line soliton,
so both line solitons experience a nontrivial transformation due to their interaction.
An example scattering event between two parallel line solitons
is presented in~\figref{fig:linelinescattering}.

\subsection{\texorpdfstring{$3 \to 3$}{3 to 3} Interactions}
Given that this 3d integrable model exhibits nontrivial scattering,
we might ask if this 3d integrable scattering exhibits factorisation,
allowing for $N \to N$ scattering events to be decomposed into a series of $2 \to 2$ interactions.
The first hint of such a property comes from the scattering transformations
that we have already computed.
Notice that the variables $\mathcal{A}$ and $\mathcal{B}$ only depend on the asymptotic data in
the scattering process, and are insensitive to any details about the interpolating bulk solution.
It is therefore reasonable to suspect that two configurations which share the same asymptotic data,
but have different bulk interpolations, might exhibit the same scattering behaviour.

\medskip

To study this question in more detail, let us consider a $3 \to 3$ scattering process
which corresponds to a $3$-soliton dressing transformation.
Applying an additional $1$-soliton transformation with parameters
$\{ \alpha_k , \hat{a}_k = (1, \hat{f}_k) \}$ to the $2$-soliton solution gives
\begin{equation}
    U_{ijk} = H_{ijk}^{-1} U_{ij} \,, \qquad
    H_{ijk}^{-1} = \id - \frac{\alpha_k - \bar\alpha_k}{z - \bar\alpha_k} \, P(a_{ijk}) \,, \qquad
    a_{ijk} = (U_{ij} \vert_{\alpha_k}) \, \hat{a}_k \,.
\end{equation}
This $3$-soliton solution is completely symmetric in its three indices,
as can be shown by a similar argument to that used to show
the symmetry of the $2$-soliton solution $U_{ij}$ in its indices.
Let us study the $1$-soliton profile associated with $\alpha_k$ far away from
the other two solitons, where $\hat{a}_i$ and $\hat{a}_j$ can be approximated by fixed values.

\medskip

Consider a region where the other two solitons can be approximated by constant maps
$\hat{f}_i = Z_i$ and $\hat{f}_j = Z_j$.
In this region, the $z$-dependent matrix $U_{ij}$ will not depend on spacetime,
and we will denote it by $U_{ij}(z; Z_i, Z_j)$.
Up to a global symmetry, the $3$-soliton solution in this region can be approximated by
a $1$-soliton solution with parameters $\alpha_k$ and
$a_{ijk} = U_{ij}(\alpha_k; Z_i, Z_j) \, \hat{a}_k$.
Therefore, if the soliton associated with $\alpha_k$ moves from a region in which
the other two solitons can be approximated by
$\{ \hat{f}_i = Z_i^{\text{in}}, \hat{f}_j = Z_j^{\text{in}} \}$ to a region in which
they are approximated by $\{ \hat{f}_i = Z_i^{\text{out}}, \hat{f}_j = Z_j^{\text{out}} \}$,
it will transform by
\begin{equation}
    a_{ijk} \mapsto U_{ij}(\alpha_k; Z_i^{\text{out}}, Z_j^{\text{out}})
    U_{ij}(\alpha_k; Z_i^{\text{in}}, Z_j^{\text{in}})^{-1} a_{ijk} \,.
\end{equation}
Let us show that this $3 \to 3$ interaction can be factorised into $2 \to 2$ interactions,
and that this factorisation is consistent because the final answer does not depend
on the order in which the solitons interact.

\medskip

The soliton associated with $\alpha_k$ will interact with the two other solitons as it moves
from the region $\{ \hat{f}_i = Z_i^{\text{in}}, \hat{f}_j = Z_j^{\text{in}} \}$
to the region $\{ \hat{f}_i = Z_i^{\text{out}}, \hat{f}_j = Z_j^{\text{out}} \}$,
and it can do this in two different orderings.
It can either interact with the soliton associated with $\alpha_i$ first
(as it passes from $Z_i^{\text{in}} \to Z_i^{\text{out}}$),
and then interact with the soliton associated with $\alpha_j$ second
(as it passes from $Z_j^{\text{in}} \to Z_j^{\text{out}}$), or the opposite.
This choice corresponds to two different factorisations of the transformation given above.
For example, the transformation above can be rewritten as
\begin{equation}\begin{aligned}\label{eq:3to3interaction}
    &
    U_{ij}(\alpha_k; Z_i^{\text{out}}, Z_j^{\text{out}})
    U_{ij}(\alpha_k; Z_i^{\text{in}}, Z_j^{\text{in}})^{-1} \\
    & =
    U_{ij}(\alpha_k; Z_i^{\text{out}}, Z_j^{\text{out}})
    U_{ij}(\alpha_k; Z_i^{\text{out}}, Z_j^{\text{in}})^{-1}
    U_{ij}(\alpha_k; Z_i^{\text{out}}, Z_j^{\text{in}})
    U_{ij}(\alpha_k; Z_i^{\text{in}}, Z_j^{\text{in}})^{-1} \\
    & =
    H_{ij}(\alpha_k; Z_i^{\text{out}}, Z_j^{\text{out}})^{-1}
    H_{ij}(\alpha_k; Z_i^{\text{out}}, Z_j^{\text{in}})
    H_{ji}(\alpha_k; Z_j^{\text{in}}, Z_i^{\text{out}})^{-1}
    H_{ji}(\alpha_k; Z_j^{\text{in}}, Z_i^{\text{in}}) \,.
\end{aligned}\end{equation}
To show that this is equivalent to a series of $2 \to 2$ interactions,
note that $H_{ij}(z; Z_i, Z_j)$ only depends on $Z_i$ and $Z_j$
through $a_{ij} = U_i(\alpha_i; Z_i) \, \hat{a}_j$.
In other words, $H_{ij}(z; Z_i, Z_j)$ is identical to the $1$-soliton solution
$U_j(z; Z_i \triangleright Z_j)^{-1}$ where $Z_i \triangleright Z_j$ denotes
the element of $\CP^1$ specified by $a_{ij} = U_i(\alpha_i; Z_i) \, \hat{a}_j$.
Therefore, the transformation~\eqref{eq:3to3interaction} can be rewritten as
\begin{equation}
    U_{j}(\alpha_k; Z_i^{\text{out}} \triangleright Z_j^{\text{out}})
    U_{j}(\alpha_k; Z_i^{\text{out}} \triangleright Z_j^{\text{in}})^{-1}
    U_{i}(\alpha_k; Z_j^{\text{in}} \triangleright Z_i^{\text{out}})
    U_{i}(\alpha_k; Z_j^{\text{in}} \triangleright Z_i^{\text{in}})^{-1} \,.
\end{equation}
This is a product of two $2 \to 2$ interactions,
and it describes a process in which the soliton associated with $\alpha_k$
first passes from $Z_i^{\text{in}} \to Z_i^{\text{out}}$ in the background $Z_j^{\text{in}}$,
and then from $Z_j^{\text{in}} \to Z_j^{\text{out}}$ in the background $Z_i^{\text{out}}$.
Alternatively, the transformation~\eqref{eq:3to3interaction} can also be factorised as
\begin{equation}
    U_{i}(\alpha_k; Z_j^{\text{out}} \triangleright Z_i^{\text{out}})
    U_{i}(\alpha_k; Z_j^{\text{out}} \triangleright Z_i^{\text{in}})^{-1}
    U_{j}(\alpha_k; Z_i^{\text{in}} \triangleright Z_j^{\text{out}})
    U_{j}(\alpha_k; Z_i^{\text{in}} \triangleright Z_j^{\text{in}})^{-1} \,.
\end{equation}
This factorisation describes the opposite order of interactions,
in which the soliton associated with $\alpha_k$ first
passes from $Z_j^{\text{in}} \to Z_j^{\text{out}}$ in the background $Z_i^{\text{in}}$,
and then from $Z_i^{\text{in}} \to Z_i^{\text{out}}$ in the background $Z_j^{\text{out}}$.
The two factorisations agree with one another as they are both rewritings of the same expression.

\begin{figure}
\centering
\begin{tikzpicture}

\newcommand{\boxHeight}{4}
\newcommand{\boxWidth}{5}
\newcommand{\boxDepth}{1}
\newcommand{\boxOffset}{0.5}
\newcommand{\lineOffset}{4}

% back of box
\coordinate (B1) at (0,0);
\draw[dashed] (B1) -- ++(-\boxOffset*\boxDepth,-\boxDepth) -- ++(0,\boxHeight) -- ++(\boxOffset*\boxDepth,\boxDepth) -- (B1);
\draw[dashed] (B1) -- ++(\boxWidth,0) -- ++(0,\boxHeight) -- ++(-\boxWidth,0);
\draw[dashed] (B1) -- ++(-\boxOffset*\boxDepth,-\boxDepth) -- ++(\boxWidth,0) -- ++(\boxOffset*\boxDepth,\boxDepth);

% red particle trajectory
\path (B1) ++(0.5-0.5*\boxOffset*\boxDepth,-0.5*\boxDepth) coordinate (P1R);
\path (P1R) ++(\lineOffset,\boxHeight) coordinate (P1RE);
\draw[thick,tab10red] (P1R) -- (P1RE);

% green line trajectory
\coordinate (L1G) at (4.5,0);
\filldraw[tab10green!40] (L1G) -- ++(-\boxOffset*\boxDepth,-\boxDepth) -- ++(-\lineOffset,\boxHeight) -- ++(\boxOffset*\boxDepth,\boxDepth) -- cycle;

% blue line trajectory
\coordinate (L1B) at (1.5,0);
\filldraw[tab10blue!40] (L1B) -- ++(-\boxOffset*\boxDepth,-\boxDepth) -- ++(0,\boxHeight) -- ++(\boxOffset*\boxDepth,\boxDepth) -- cycle;

% red particle foreground
\path (P1R) ++(0.25*\lineOffset,0.25*\boxHeight) coordinate (P1RB);
\node[rotate=20,tab10red] at (P1RB) {$\times$};
\draw[thick,tab10red] (P1RB) -- (P1RE);

% green line foreground
\filldraw[tab10green!40] (L1G) -- ++(-\boxOffset*\boxDepth,-\boxDepth) -- ++(-0.75*\lineOffset,0.75*\boxHeight) -- ++(\boxOffset*\boxDepth,\boxDepth) -- cycle;

% red particle foreground-foreground
\path (P1R) ++(0.5*\lineOffset,0.5*\boxHeight) coordinate (P1RG);
\node[rotate=20,tab10red] at (P1RG) {$\times$};
\draw[thick,tab10red] (P1RG) -- (P1RE);

% dashed trajectory outlines
\draw[dashed,tab10blue!40] (L1B) -- ++(-\boxOffset*\boxDepth,-\boxDepth) -- ++(0,\boxHeight) -- ++(\boxOffset*\boxDepth,\boxDepth) -- (L1B);
\draw[dashed,tab10green!40] (L1G) -- ++(-\boxOffset*\boxDepth,-\boxDepth) -- ++(-\lineOffset,\boxHeight) -- ++(\boxOffset*\boxDepth,\boxDepth) -- (L1G);
\draw[dashed,tab10red] (P1R) -- ++(\lineOffset,\boxHeight);

% front of box
\path (B1) ++(\boxWidth-\boxOffset*\boxDepth,\boxHeight-\boxDepth) coordinate (F1);
\draw[dashed] (F1) -- ++(\boxOffset*\boxDepth,\boxDepth);
\draw[dashed] (F1) -- ++(-\boxWidth,0);
\draw[dashed] (F1) -- ++(0,-\boxHeight);

% equals sign
\node[scale=2] at (0.5*\boxWidth+0.5*8-0.5*\boxOffset*\boxDepth,0.5*\boxHeight-0.5*\boxDepth) {$=$};

% back of box
\coordinate (B2) at (8,0);
\draw[dashed] (B2) -- ++(-\boxOffset*\boxDepth,-\boxDepth) -- ++(0,\boxHeight) -- ++(\boxOffset*\boxDepth,\boxDepth) -- (B2);
\draw[dashed] (B2) -- ++(\boxWidth,0) -- ++(0,\boxHeight) -- ++(-\boxWidth,0);
\draw[dashed] (B2) -- ++(-\boxOffset*\boxDepth,-\boxDepth) -- ++(\boxWidth,0) -- ++(\boxOffset*\boxDepth,\boxDepth);

% red particle trajectory
\path (B2) ++(0.5-0.5*\boxOffset*\boxDepth,-0.5*\boxDepth) coordinate (P2R);
\path (P2R) ++(\lineOffset,\boxHeight) coordinate (P2RE);
\draw[thick,tab10red] (P2R) -- (P2RE);

% green line trajectory
\coordinate (L2G) at (12.5,0);
\filldraw[tab10green!40] (L2G) -- ++(-\boxOffset*\boxDepth,-\boxDepth) -- ++(-\lineOffset,\boxHeight) -- ++(\boxOffset*\boxDepth,\boxDepth) -- cycle;

% red particle foreground
\path (P2R) ++(0.5*\lineOffset,0.5*\boxHeight) coordinate (P2RG);
\draw[thick,tab10red] (P2RG) -- (P2RE);

% blue line trajectory
\coordinate (L2B) at (11.5,0);
\filldraw[tab10blue!40] (L2B) -- ++(-\boxOffset*\boxDepth,-\boxDepth) -- ++(0,\boxHeight) -- ++(\boxOffset*\boxDepth,\boxDepth) -- cycle;

% green line foreground
\filldraw[tab10green!40] (L2G) -- ++(-\boxOffset*\boxDepth,-\boxDepth) -- ++(-0.25*\lineOffset,0.25*\boxHeight) -- ++(\boxOffset*\boxDepth,\boxDepth) -- cycle;

% red particle foreground-foreground
\path (P2R) ++(0.75*\lineOffset,0.75*\boxHeight) coordinate (P2RB);
\draw[thick,tab10red] (P2RB) -- (P2RE);

% interaction points
\node[rotate=20,tab10red] at (P2RG) {$\times$};
\node[rotate=20,tab10red] at (P2RB) {$\times$};

% dashed trajectory outlines
\draw[dashed,tab10blue!40] (L2B) -- ++(-\boxOffset*\boxDepth,-\boxDepth) -- ++(0,\boxHeight) -- ++(\boxOffset*\boxDepth,\boxDepth) -- (L2B);
\draw[dashed,tab10green!40] (L2G) -- ++(-\boxOffset*\boxDepth,-\boxDepth) -- ++(-\lineOffset,\boxHeight) -- ++(\boxOffset*\boxDepth,\boxDepth) -- (L2G);
\draw[dashed,tab10red] (P2R) -- ++(\lineOffset,\boxHeight);

% front of box
\path (B2) ++(\boxWidth-\boxOffset*\boxDepth,\boxHeight-\boxDepth) coordinate (F2);
\draw[dashed] (F2) -- ++(\boxOffset*\boxDepth,\boxDepth);
\draw[dashed] (F2) -- ++(-\boxWidth,0);
\draw[dashed] (F2) -- ++(0,-\boxHeight);

\end{tikzpicture}
\caption{
    Scattering between lump solitons and line solitons obeys a nontrivial constraint,
    leading to the equivalence of these two scattering processes.
    While the intermediate profiles may be different,
    the asymptotic profiles are the same on both sides.
}
\label{fig:lumplineline}
\end{figure}
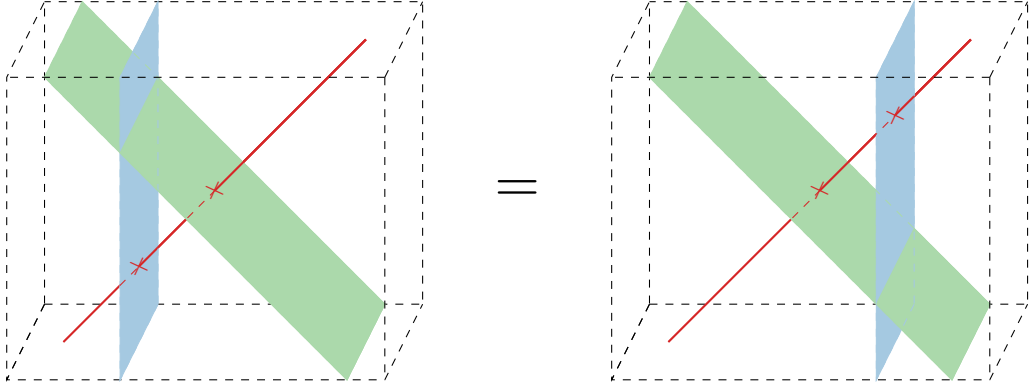

\begin{figure}
\centering
\begin{tikzpicture}

\newcommand{\boxHeight}{4}
\newcommand{\boxWidth}{5}
\newcommand{\boxDepth}{1}
\newcommand{\boxOffset}{0.5}
\newcommand{\lineOffset}{4}

% back of box
\coordinate (B1) at (0,0);
\draw[dashed] (B1) -- ++(-\boxOffset*\boxDepth,-\boxDepth) -- ++(0,\boxHeight) -- ++(\boxOffset*\boxDepth,\boxDepth) -- (B1);
\draw[dashed] (B1) -- ++(\boxWidth,0) -- ++(0,\boxHeight) -- ++(-\boxWidth,0);
\draw[dashed] (B1) -- ++(-\boxOffset*\boxDepth,-\boxDepth) -- ++(\boxWidth,0) -- ++(\boxOffset*\boxDepth,\boxDepth);

% red line trajectory
\coordinate (L1R) at (0.5,0);
\filldraw[tab10red!40] (L1R) -- ++(-\boxOffset*\boxDepth,-\boxDepth) -- ++(\lineOffset,\boxHeight) -- ++(\boxOffset*\boxDepth,\boxDepth) -- cycle;

% green line trajectory
\coordinate (L1G) at (4.5,0);
\filldraw[tab10green!40] (L1G) -- ++(-\boxOffset*\boxDepth,-\boxDepth) -- ++(-\lineOffset,\boxHeight) -- ++(\boxOffset*\boxDepth,\boxDepth) -- cycle;

% blue line trajectory
\coordinate (L1B) at (1.5,0);
\filldraw[tab10blue!40] (L1B) -- ++(-\boxOffset*\boxDepth,-\boxDepth) -- ++(0,\boxHeight) -- ++(\boxOffset*\boxDepth,\boxDepth) -- cycle;

% red line foreground
\path (L1R) ++(0.25*\lineOffset,0.25*\boxHeight) coordinate (L1RB);
\filldraw[tab10red!40] (L1RB) -- ++(-\boxOffset*\boxDepth,-\boxDepth) -- ++(0.75*\lineOffset,0.75*\boxHeight) -- ++(\boxOffset*\boxDepth,\boxDepth) -- cycle;

% green line foreground
\filldraw[tab10green!40] (L1G) -- ++(-\boxOffset*\boxDepth,-\boxDepth) -- ++(-0.75*\lineOffset,0.75*\boxHeight) -- ++(\boxOffset*\boxDepth,\boxDepth) -- cycle;

% red line foreground-foreground
\path (L1R) ++(0.5*\lineOffset,0.5*\boxHeight) coordinate (L1RG);
\filldraw[tab10red!40] (L1RG) -- ++(-\boxOffset*\boxDepth,-\boxDepth) -- ++(0.5*\lineOffset,0.5*\boxHeight) -- ++(\boxOffset*\boxDepth,\boxDepth) -- cycle;

% dashed trajectory outlines
\draw[dashed,tab10blue!40] (L1B) -- ++(-\boxOffset*\boxDepth,-\boxDepth) -- ++(0,\boxHeight) -- ++(\boxOffset*\boxDepth,\boxDepth) -- (L1B);
\draw[dashed,tab10green!40] (L1G) -- ++(-\boxOffset*\boxDepth,-\boxDepth) -- ++(-\lineOffset,\boxHeight) -- ++(\boxOffset*\boxDepth,\boxDepth) -- (L1G);
\draw[dashed,tab10red!40] (L1R) -- ++(-\boxOffset*\boxDepth,-\boxDepth) -- ++(\lineOffset,\boxHeight) -- ++(\boxOffset*\boxDepth,\boxDepth) -- cycle;

% front of box
\path (B1) ++(\boxWidth-\boxOffset*\boxDepth,\boxHeight-\boxDepth) coordinate (F1);
\draw[dashed] (F1) -- ++(\boxOffset*\boxDepth,\boxDepth);
\draw[dashed] (F1) -- ++(-\boxWidth,0);
\draw[dashed] (F1) -- ++(0,-\boxHeight);

% equals sign
\node[scale=2] at (0.5*\boxWidth+0.5*8-0.5*\boxOffset*\boxDepth,0.5*\boxHeight-0.5*\boxDepth) {$=$};

% back of box
\coordinate (B2) at (8,0);
\draw[dashed] (B2) -- ++(-\boxOffset*\boxDepth,-\boxDepth) -- ++(0,\boxHeight) -- ++(\boxOffset*\boxDepth,\boxDepth) -- (B2);
\draw[dashed] (B2) -- ++(\boxWidth,0) -- ++(0,\boxHeight) -- ++(-\boxWidth,0);
\draw[dashed] (B2) -- ++(-\boxOffset*\boxDepth,-\boxDepth) -- ++(\boxWidth,0) -- ++(\boxOffset*\boxDepth,\boxDepth);

% red line trajectory
\coordinate (L2R) at (8.5,0);
\filldraw[tab10red!40] (L2R) -- ++(-\boxOffset*\boxDepth,-\boxDepth) -- ++(\lineOffset,\boxHeight) -- ++(\boxOffset*\boxDepth,\boxDepth) -- cycle;

% green line trajectory
\coordinate (L2G) at (12.5,0);
\filldraw[tab10green!40] (L2G) -- ++(-\boxOffset*\boxDepth,-\boxDepth) -- ++(-\lineOffset,\boxHeight) -- ++(\boxOffset*\boxDepth,\boxDepth) -- cycle;

% red line foreground
\path (L2R) ++(0.5*\lineOffset,0.5*\boxHeight) coordinate (L2RG);
\filldraw[tab10red!40] (L2RG) -- ++(-\boxOffset*\boxDepth,-\boxDepth) -- ++(0.5*\lineOffset,0.5*\boxHeight) -- ++(\boxOffset*\boxDepth,\boxDepth) -- cycle;

% blue line trajectory
\coordinate (L2B) at (11.5,0);
\filldraw[tab10blue!40] (L2B) -- ++(-\boxOffset*\boxDepth,-\boxDepth) -- ++(0,\boxHeight) -- ++(\boxOffset*\boxDepth,\boxDepth) -- cycle;

% green line foreground
\filldraw[tab10green!40] (L2G) -- ++(-\boxOffset*\boxDepth,-\boxDepth) -- ++(-0.25*\lineOffset,0.25*\boxHeight) -- ++(\boxOffset*\boxDepth,\boxDepth) -- cycle;

% red line foreground-foreground
\path (L2R) ++(0.75*\lineOffset,0.75*\boxHeight) coordinate (L2RB);
\filldraw[tab10red!40] (L2RB) -- ++(-\boxOffset*\boxDepth,-\boxDepth) -- ++(0.25*\lineOffset,0.25*\boxHeight) -- ++(\boxOffset*\boxDepth,\boxDepth) -- cycle;

% dashed trajectory outlines
\draw[dashed,tab10blue!40] (L2B) -- ++(-\boxOffset*\boxDepth,-\boxDepth) -- ++(0,\boxHeight) -- ++(\boxOffset*\boxDepth,\boxDepth) -- cycle;
\draw[dashed,tab10green!40] (L2G) -- ++(-\boxOffset*\boxDepth,-\boxDepth) -- ++(-\lineOffset,\boxHeight) -- ++(\boxOffset*\boxDepth,\boxDepth) -- cycle;
\draw[dashed,tab10red!40] (L2R) -- ++(-\boxOffset*\boxDepth,-\boxDepth) -- ++(\lineOffset,\boxHeight) -- ++(\boxOffset*\boxDepth,\boxDepth) -- cycle;

% front of box
\path (B2) ++(\boxWidth-\boxOffset*\boxDepth,\boxHeight-\boxDepth) coordinate (F2);
\draw[dashed] (F2) -- ++(\boxOffset*\boxDepth,\boxDepth);
\draw[dashed] (F2) -- ++(-\boxWidth,0);
\draw[dashed] (F2) -- ++(0,-\boxHeight);

\end{tikzpicture}
\caption{
    Scattering between parallel line solitons obeys a 3d analogue of the 2d Yang-Baxter equation.
    After compactifying the theory and reducing to two dimensions,
    parallel line soliton scattering in 3d corresponds to 2d lump soliton scattering.
}
\label{fig:3parallellines}
\end{figure}
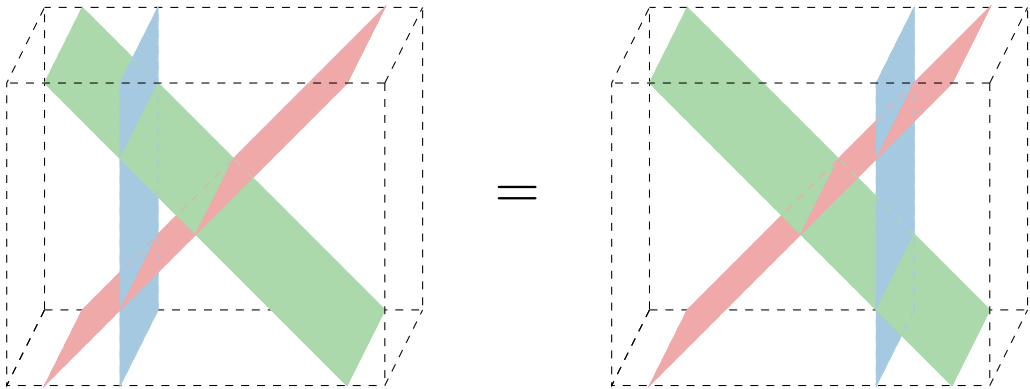

\medskip

Let us consider some examples in which these statements can be given a physical interpretation.
Since lumps do not scatter with one another, at least two of the solitons must be line solitons
in order for these statements to be nontrivial.
The simplest case is a $3 \to 3$ scattering event between one lump soliton and two line solitons.
Following the lump soliton as it moves through space, it will generically intersect
each of the line solitons at some finite time.
There are two inequivalent orderings in which these interactions may take place,
as depicted in~\figref{fig:lumplineline}.
Each factorisation of the $3 \to 3$ scattering transformation represents one of these orderings,
and their agreement implies that the final profile of the lump soliton
does not depend on the path taken.

\medskip

An analogous statement applies to $3 \to 3$ scattering between parallel lines,
as depicted in~\figref{fig:3parallellines}.
In this case, the intermediate profiles of all three lines will differ between the two orderings,
but the final profiles will nonetheless agree, independently of the path taken.
This can also be related to a classical analogue of the 2d Yang-Baxter equation.
If a 2d integrable theory is embedded into three dimensions,
the 2d lump solitons appear as line solitons in the 3d theory.
Therefore, 2d lump soliton scattering corresponds to 3d parallel line soliton scattering,
and the relationship depicted in~\figref{fig:3parallellines}
corresponds to a classical analogue of the 2d Yang-Baxter equation.

\begin{figure}
\centering
\begin{tikzpicture}

\newcommand{\squareLength}{4}
\newcommand{\squareSpacing}{1}

\newcommand{\boxWidth}{0.15}
\newcommand{\boxHeight}{0.5}
\newcommand{\boxOffset}{0.75}

% first square
\coordinate (S1) at (0,0);
\draw[dashed] (S1) -- ++(\squareLength,0) -- ++(0,\squareLength) -- ++(-\squareLength,0) -- (S1);
\node at (0.5*\squareLength,\squareLength+0.4) {$t = -1$};

\newcommand{\paramTime}{-1}

% red line
\path (S1) ++(0.6-0.4*\paramTime,0) coordinate (R1);
\draw[thick,tab10red] (R1) -- ++(\squareLength-1.2,\squareLength);

% green line
\path (S1) ++(0.6-0.4*\paramTime,\squareLength) coordinate (G1);
\draw[thick,tab10green] (G1) -- ++(\squareLength-1.2,-\squareLength);

% blue line
\path (S1) ++(2+1.25*\paramTime,0) coordinate (B1);
\draw[thick,tab10blue] (B1) -- ++(0,\squareLength);

% section boxes
\path (B1) ++(-0.5*\boxWidth,\boxOffset) coordinate (bL1);
\draw[black!80] (bL1) -- ++(\boxWidth,0) -- ++(0,\boxHeight) -- ++(-\boxWidth,0) -- (bL1);
\path (B1) ++(-0.5*\boxWidth,\squareLength-\boxHeight-\boxOffset) coordinate (bU1);
\draw[black!80] (bU1) -- ++(\boxWidth,0) -- ++(0,\boxHeight) -- ++(-\boxWidth,0) -- (bU1);

% second square
\path (S1) ++(\squareLength+\squareSpacing,0) coordinate (S2);
\draw[dashed] (S2) -- ++(\squareLength,0) -- ++(0,\squareLength) -- ++(-\squareLength,0) -- (S2);
\node at (1.5*\squareLength+\squareSpacing,\squareLength+0.4) {$t = -0.1$};

\renewcommand{\paramTime}{-0.1}

% red line
\path (S2) ++(0.6-0.4*\paramTime,0) coordinate (R2);
\draw[thick,tab10red] (R2) -- ++(\squareLength-1.2,\squareLength);

% green line
\path (S2) ++(0.6-0.4*\paramTime,\squareLength) coordinate (G2);
\draw[thick,tab10green] (G2) -- ++(\squareLength-1.2,-\squareLength);

% blue line
\path (S2) ++(2+1.25*\paramTime,0) coordinate (B2);
\draw[thick,tab10blue] (B2) -- ++(0,\squareLength);

% section boxes
\path (B2) ++(-0.5*\boxWidth,\boxOffset) coordinate (bL2);
\draw[black!80] (bL2) -- ++(\boxWidth,0) -- ++(0,\boxHeight) -- ++(-\boxWidth,0) -- (bL2);
\path (B2) ++(-0.5*\boxWidth,\squareLength-\boxHeight-\boxOffset) coordinate (bU2);
\draw[black!80] (bU2) -- ++(\boxWidth,0) -- ++(0,\boxHeight) -- ++(-\boxWidth,0) -- (bU2);

% third square
\path (S2) ++(\squareLength+\squareSpacing,0) coordinate (S3);
\draw[dashed] (S3) -- ++(\squareLength,0) -- ++(0,\squareLength) -- ++(-\squareLength,0) -- (S3);
\node at (2.5*\squareLength+2*\squareSpacing,\squareLength+0.4) {$t = +1$};

\renewcommand{\paramTime}{1}

% red line
\path (S3) ++(0.6-0.4*\paramTime,0) coordinate (R3);
\draw[thick,tab10red] (R3) -- ++(\squareLength-1.2,\squareLength);

% green line
\path (S3) ++(0.6-0.4*\paramTime,\squareLength) coordinate (G3);
\draw[thick,tab10green] (G3) -- ++(\squareLength-1.2,-\squareLength);

% blue line
\path (S3) ++(2+1.25*\paramTime,0) coordinate (B3);
\draw[thick,tab10blue] (B3) -- ++(0,\squareLength);

% section boxes
\path (B3) ++(-0.5*\boxWidth,\boxOffset) coordinate (bL3);
\draw[black!80] (bL3) -- ++(\boxWidth,0) -- ++(0,\boxHeight) -- ++(-\boxWidth,0) -- (bL3);
\path (B3) ++(-0.5*\boxWidth,\squareLength-\boxHeight-\boxOffset) coordinate (bU3);
\draw[black!80] (bU3) -- ++(\boxWidth,0) -- ++(0,\boxHeight) -- ++(-\boxWidth,0) -- (bU3);

\end{tikzpicture}
\caption{
    Generically, three line solitons are required to have a scattering event.
    The boxed segments of the blue line soliton are connected at asymptotic times,
    but they move into different regions at intermediate times.
    The profiles of these segments match before and after scattering,
    despite the fact that they have taken different paths.
}
\label{fig:3generallines}
\end{figure}
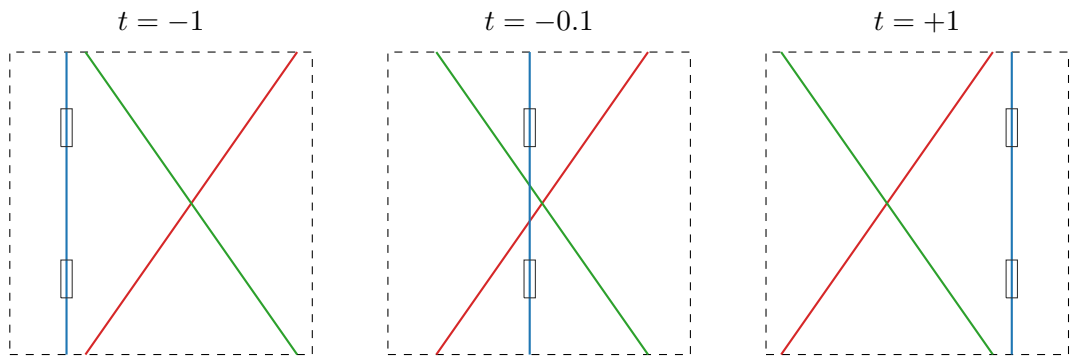

\medskip

Finally, a generic scattering event between three line solitons
is depicted in~\figref{fig:3generallines}.
There are segments of each line soliton which are connected at asymptotic times,
but which are disconnected for some intermediate times.
These segments interact with the other two line solitons in different orders,
meaning that their intermediate profiles may disagree,
but their profiles will match when they reconnect after both interactions.

\medskip

Following the argument presented above,
a general $N \to N$ interaction can be factorised into a series of $2 \to 2$ interactions,
and the final result does not depend on the order in which the interactions take place.
When applied to $4 \to 4$ scattering between line solitons,
this statement implies an equivalence between different orders of interactions,
resembling a classical analogue of the Zamolodchikov tetrahedron equation~\cite{Zam80,Zam81}.
This should be regarded as a classical consistency condition with an analogous form,
rather than a direct derivation of the quantum tetrahedron equation.

\section{Relation to 2d Integrability}\label{sec:reduction}
The 3d integrable chiral model is related to certain 2d integrable theories
by a one-dimensional reduction.
Consider the 3d integrable chiral model with $\grp{G} = \grp{SU}(2)$,
on the manifold $\fR^{1,1} \times S^1$ where we have compactified the $y$-direction.
We will continue to denote the $S^1$ coordinate by $y$, but it will be subject to the relation
\begin{equation}
    y \sim y + 2 \pi R \,.
\end{equation}
The radius will eventually be taken to zero ($R \to 0$) to recover a 2d integrable theory.
Along the compact direction, we will impose twisted boundary conditions given by
\begin{equation}
    g(y + 2 \pi R) = \Omega \, g(y) \, \Omega^{-1} \,, \qquad
    \Omega = \exp (2 \pi R \, \mu \, \mathsf{t}_1) \,.
\end{equation}
This expression is written in terms of a basis of $\alg{su}(2)$ which is given by
\begin{equation}
    \mathsf{t}_1 = \frac{1}{2}
    \begin{pmatrix}
        \iu  & 0    \\
        0    & -\iu \\
    \end{pmatrix} \,, \qquad
    \mathsf{t}_2 = \frac{1}{2}
    \begin{pmatrix}
        0    & 1    \\
        -1   & 0    \\
    \end{pmatrix} \,, \qquad
    \mathsf{t}_3 = \frac{1}{2}
    \begin{pmatrix}
        0    & \iu  \\
        \iu  & 0    \\
    \end{pmatrix} \,.
\end{equation}
This basis satisfies the commutation relations
$[ \mathsf{t}_i , \mathsf{t}_j ] = \varepsilon_{ijk} \mathsf{t}_k$ with $\varepsilon_{123} = +1$,
and it is orthogonal with respect to the matrix trace
$\tr (\mathsf{t}_i \mathsf{t}_j) = -\frac{1}{2} \delta_{ij}$.
The parameter $\mu$ appearing in $\Omega$ controls the amount of twisting
in the boundary conditions, and the choice $\mu = 0$ corresponds to periodic boundary conditions.

\medskip

Introducing twisted boundary conditions is equivalent to coupling the theory
to a background gauge field along the compact direction.
We can see this by rewriting the fundamental field as
\begin{equation}
    g = \omega \, h \, \omega^{-1} \,, \qquad
    \omega = \exp (\mu \, y \, \mathsf{t}_1) \,.
\end{equation}
The fixed configuration $\omega$ accounts for all of the twisting in the boundary condition,
such that the field $h$ is periodic in the $y$-direction.
Then, partial derivatives on $g$ can be written as covariant derivatives on $h$ via the relation
\begin{equation}
    \pd_\mu g = \omega (\Dr_\mu h) \omega^{-1} \,, \qquad
    \Dr_\mu = \pd_\mu h + \omega^{-1} \pd_\mu \omega \, h - h \, \omega^{-1} \pd_\mu \omega \,.
\end{equation}
This is the covariant derivative for a background gauge field $A_\mu = \omega^{-1} \pd_\mu \omega$
acting via the diagonal $\grp{G}$-action.
In the present example, the only nontrivial component of the background gauge field is
$A_y = \mu \, \mathsf{t}_1$.

\medskip

Given that $h$ is periodic in the $y$-direction, we can perform a Fourier expansion to trade this
single $y$-dependent field for an infinite set of $y$-independent modes.
Almost all of these modes will have a mass that tends to infinity in the limit $R \to 0$,
meaning that they can be neglected in the low energy effective theory.
To determine the masses of the various modes, we expand the action around
a background configuration and keep terms up to quadratic order in the fluctuations.
Let us adopt the following parameterisation of $\grp{SU}(2)$, which matches the one given earlier.
\begin{equation}\begin{aligned}
    &
    \begin{aligned}
        X_1 & = \cos(\phi) \\
        X_2 & = \sin(\phi) \, \frac{1 - \vert Z \vert^2}{1 + \vert Z \vert^2}
    \end{aligned}
    \quad & \quad &
    h =
    \begin{pmatrix}
        X_1 + \iu X_2   & X_3 + \iu X_4 \\
        -X_3 + \iu X_4  & X_1 - \iu X_2
    \end{pmatrix}
    \\
    & X_3 = \sin(\phi) \, \frac{-\iu (Z - \bar{Z})}{1 + \vert Z \vert^2}
    \quad & \quad &
    \qquad X_1^2 + X_2^2 + X_3^2 + X_4^2 = 1 \\
    & X_4 = \sin(\phi) \, \frac{Z + \bar{Z}}{1 + \vert Z \vert^2}
    \quad & \quad &
    \dr s^2 = \dr \phi^2 + 4 \sin^2 (\phi) \frac{\dr Z \dr \bar{Z}}{(1 + \vert Z \vert^2)^2}
\end{aligned}\end{equation}
The quadratic order terms in the fluctuations will only come from the kinetic term in the action,
and the kinetic term with the background gauge field is given by
\begin{equation}
    -\frac{1}{2} \tr (h^{-1} \Dr_\mu h h^{-1} \Dr^\mu h) =
    \pd_\mu \phi \pd^\mu \phi + 4 \sin^2 (\phi) \frac{\vert \Dr_\mu Z \vert^2}{(1 + \vert Z \vert^2)^2} \,.
\end{equation}
The derivatives acting on $\phi$ are not modified by the background gauge field,
and the covariant derivative acting on $Z$ has only one nontrivial component
which is given by $\Dr_y Z = \pd_y Z - \iu \mu Z$.
Let us expand the fields around the constant background configuration
$(\phi = \frac{\pi}{2} , Z = 0)$, and write the fluctuations as
$\phi = \frac{\pi}{2} + \delta \phi$ and $Z = \delta Z$.
To quadratic order in the fluctuations, the action density is given by
\begin{equation}
    \pd_\mu (\delta \phi) \pd^\mu (\delta \phi) + 4 \, \vert \Dr_\mu (\delta Z) \vert^2 \,.
\end{equation}
Expanding the variations into Fourier modes as
$\delta \phi = \sum \phi_n e^{\iu n y / R}$ and $\delta Z = \sum Z_n e^{\iu n y / R}$,
the mass of each mode can be determined by studying the quadratic action density.
Substituting in the expansions and integrating over the $S^1$ parameterised by $y$,
the quadratic action density becomes
\begin{equation}
    \sum_n \bigg(
        \pd_\mu \phi_n \pd^\mu \phi_{-n} + \frac{n^2}{R^2} \, \phi_n \phi_{-n}
        + 4 \, \vert \pd_\mu Z_n \vert^2 + 4 \bigg( \frac{n}{R} - \mu \bigg)^{\! 2} \vert Z_n \vert^2
    \bigg) \,.
\end{equation}
The twisted boundary conditions have shifted the masses of the $Z_n$ modes by $\mu$,
meaning that the zero mode is no longer massless.
Nonetheless, if we take the limit $R \to 0$ while keeping all other parameters fixed,
the lowest mass modes are still $\phi_0$ and $Z_0$
with the masses of all other modes tending to infinity.
From this analysis, we expect the associated 2d theory to be a theory of
one massless scalar and two massive scalars.
Notice that, while $h$ will be constant in $y$ in this zero mode sector,
the field $g$ will have nontrivial $y$-dependence due to the twisted boundary conditions.

\medskip

To compute the effective action for the zero modes,
we declare that $h$ is constant in the $y$-direction and compute the integral over $S^1$.
Up to an overall factor of $\vol(S^1) = 2 \pi R$, this gives the 2d theory
\begin{equation}\begin{gathered}
    S_{\text{2d}}[h] = S_{\text{WZW}}[h]
    + \mu^2 \int_{\fR^2} \! \dr^2 x \, \tr (h^{-1} \mathsf{t}_1 h \mathsf{t}_1) \,, \\
    S_{\text{WZW}}[h] = 
    \frac{1}{2} \int_{\fR^2} \! \tr ( h^{-1} \dr h \wedge \star h^{-1} \dr h )
    + \frac{1}{3} \int_{\fR^2 \times [0,1]} \hspace{-2.2em}
    \tr (\tilde{h}^{-1} \dr \tilde{h} \wedge \tilde{h}^{-1} \dr \tilde{h} \wedge \tilde{h}^{-1} \dr \tilde{h}) \,.
\end{gathered}\end{equation}
This deformation of the 2d WZW model generates a mass for two of the scalar fields.
The twisted boundary conditions in 3d have led to this mass deformation term in 2d,
both of which break the symmetry group from $\grp{SU}(2)$ down to $\grp{U}(1)$.

\medskip

In order to connect to another 2d model, let us consider the ansatz
$\phi = \frac{\pi}{2}$ and $Z = \tan (\frac{\varphi}{4})$.
This reduces the problem from a system of three scalar fields to an equation for
a single real scalar field $\varphi(t, x)$.
Substituting this ansatz into the 2d equation of motion gives
\begin{equation}
    \pd_t^2 \varphi - \pd_x^2 \varphi + \mu^2 \sin (\varphi) = 0 \,.
\end{equation}
This is the equation of motion of the 2d sine-Gordon model,
which admits lump soliton solutions of the form
\begin{equation}
    \varphi(t, x) = 4 \arctan \exp (\mu \gamma (x - v t) + \delta) \,.
\end{equation}
In this expression, $\gamma = 1 / \sqrt{1 - v^2}$ is the Lorentz factor and $v$ is the velocity.
These 2d sine-Gordon solitons can be lifted to solutions of the 3d integrable chiral model,
and these solutions can be identified with line solitons
whose energy density is constant along the $y$-direction~\cite{Lee89}.
In terms of the parameters of the dressing transformations, they correspond to the choices
\begin{equation}
    \alpha = \iu \, r \,, \qquad
    f(w) = \exp (-\tfrac{\mu}{r} w + \delta) \,.
\end{equation}
The velocity of these lines points entirely along the $x$-direction,
normal to the profile of the lines,
so the effective velocity is equal to the lump velocity $v = \frac{-1 + r^2}{1 + r^2}$.
Interpreting these lines as domain walls in the 3d theory,
they all interpolate between the poles $Z = 0$ and $Z = \infty$.

\medskip

It is possible to extract the time delay for lump soliton scattering in the 2d sine-Gordon model
from line soliton scattering in the 3d integrable chiral model.
Consider a scattering event between two line solitons defined by
$\{ \alpha_i = \iu r_i , \hat{f}_i(w) = \exp (-\frac{\mu}{r_i} w + \delta_i) \}$.
In this case, the net effect of 3d line soliton scattering can be expressed as
\begin{equation}
    f \mapsto \mathcal{A} \, f \,, \qquad
    \mathcal{A} = \frac{
        \vert \alpha_1 - \bar\alpha_2 \vert^2
    }{
        \vert \alpha_1 - \alpha_2 \vert^2
    } \,.
\end{equation}
This can be absorbed into a shift $\Delta \delta = \log(\mathcal{A})$,
which corresponds to a real time delay.
To compare with the literature, we can take the centre of mass frame
in which $v_1 = -v_2$ (which corresponds to $r_2 = 1/r_1$).
Then, we can rewrite the scattering transformation as $\mathcal{A} = 1 / v_1^2$,
and the shift becomes $\Delta \delta = -2 \log(v_1)$, which matches the known result.

\section{Outlook}
The purpose of this paper has been to describe classical soliton scattering
in the 2+1d integrable chiral model as a case study for higher-dimensional integrable scattering.
Generic lump solitons do not scatter off each other~\cite{MZ81,War88},
but extended line solitons interact nontrivially
with lump solitons and with other line solitons~\cite{Lee89}.
Equation~\eqref{eq:scatteringtransformation} presents the scattering transformation
applied to a soliton as it interacts with a generic line soliton.
The effect of this scattering on lump solitons is a rescaling of the height, a time delay,
and a shift normal to the trajectory of the soliton.
On line solitons, it corresponds to a time delay, and to a shift in the asymptotic values.

\medskip

Furthermore, $3 \to 3$ interactions factorise into a series of $2 \to 2$ interactions,
and this factorisation is consistent because it does not depend on the order of interactions.
This equality is depicted for a lump soliton
interacting with two line solitons in~\figref{fig:lumplineline},
and for an interaction between three parallel line solitons in~\figref{fig:3parallellines}.
Moreover, the statement of factorisation applies to $N \to N$ scattering events,
which can be decomposed into a series of $2 \to 2$ interactions.
Applied to $4 \to 4$ line soliton scattering,
this resembles a classical analogue of the Zamolodchikov tetrahedron equation~\cite{Zam80,Zam81}.

\medskip

Looking towards the future, more general line solitons and their scattering
could be studied in the 2+1d integrable chiral model.
The line solitons studied in this paper were generated by the ansatz
\begin{equation}
    f(w) = \exp \big( c \, (e^{\iu \psi} w - \rchi_0) \big) + Z_0 \,.
\end{equation}
More generally, the function $f(w)$ could involve
sums and ratios of exponential or polynomial factors,
and these will correspond to more complicated soliton configurations.
The space of lump solitons is split into different sectors by the degree of $f(w)$,
and each of these sectors is described by a finite set of parameters.
It would be valuable to determine analogous statements about the space of line solitons,
or at least to find a conserved quantity which usefully organises these solutions.
One approach might start by comparing the soliton data used in this paper
with the geometric data studied in~\cite{Mas05}.

\medskip

In addition, the 2+1d integrable chiral model has recently been shown to arise
as a decoupling limit of the membrane worldvolume theory~\cite{Ost26}.
If the solitons of the 3d integrable chiral model admit a sensible interpretation in that theory,
then it is also worth asking if soliton scattering has a meaningful interpretation.

\medskip

Moving to the quantum theory, some objectives would be
to investigate quantum integrability of the 3d ICM,
and to compute its renormalization group flow.
One approach to investigating quantum integrability would be
to study the anomalies of the 5d CS description,
following the approach presented in~\cite{Cos21,BSS23}.
In this context, anomalies of the holomorphic-topological gauge theory
are interpreted as obstructions to quantum integrability on spacetime.
If the theory is quantum integrable,
or if it can be made quantum integrable by coupling to additional fields,
then its renormalization group flow should determine whether or not
classical soliton scattering is a good approximation for quantum scattering.

\medskip

Given a 2+1d integrable quantum field theory,
it may be possible to compare classical soliton scattering
to a perturbative scattering computation.
The soliton solutions could be treated as a background,
and scattering of the quantum fluctuations could be studied
using the collective coordinate method~\cite{GJS76}.
In this manner, it may be possible to circumvent the problem
of defining an asymptotic state related to the line solitons.
More ambitiously, an integrability bootstrap program for higher-dimensional integrable systems
may yield the exact scattering amplitudes for these quantum field theories.

\medskip

Higher-dimensional integrable systems have also appeared in recent top-down holographic dualities
between 4d theories on asymptotically flat spacetimes
and 2d chiral algebras~\cite{CPS23a,CPS23b,BCZ26}.
In these proposals, the 4d theory is integrable and can be described by
a 6d holomorphic-topological theory using the machinery of twistor theory.
These 4d theories exhibit nontrivial scattering between extended solitons,
just like the scattering studied in this paper,
and the interpretation of this scattering in the holographic context is worth investigating.

\section*{Acknowledgements}
It is a pleasure to thank
Tim Adamo,
Roland Bittleston,
Kevin Costello,
Ryan Cullinan,
Ben Hoare,
Joaquin Liniado,
Lionel Mason,
Tristan McLoughlin,
Vera Posch,
Bernd Schroers,
Beno{\^i}t Vicedo,
and Masahito Yamazaki
for illuminating discussions.
This work benefited from presentations at
`Integrability and Nonequilibrium Phenomena in Spacetime-Modulated Systems',
`Higher-dimensional integrability and holography',
and `Integrability, Dualities and Deformations 2026',
where feedback from the organisers and participants was greatly appreciated.
The author was supported by an ERC Consolidator/UKRI Frontier grant (TwistorQFT EP/Z000157/1).

\printbibliography

\end{document}